
%
%
\input phyzzx
\tolerance=1000
\sequentialequations
\def\rl{\rightline}

\def\t1{{\tilde 1}}

\def\AEF{Alon E. Faraggi}
\def\DVN{D. V. Nanopoulos}

\def\SSM{supersymmetric standard model}
\def\NPB#1#2#3{Nucl. Phys. B {\bf#1} (19#2) #3}
\def\PLB#1#2#3{Phys. Lett. B {\bf#1} (19#2) #3}
\def\PRD#1#2#3{Phys. Rev. D {\bf#1} (19#2) #3}

\def\MODA#1#2#3{Mod. Phys. Lett. A {\bf#1} (19#2) #3}

\REF\EKN{J. Ellis, S. Kelley and \DVN, CERN--TH.6140/91;
P. Langacker, University of Pennsylvania preprint UPR--0435T (1990);
U. Amaldi, W. de Boer and F. F\"urstenau, \PLB{260}{91}{447}.}
\REF\GSW{M. Green, J. Schwarz and E. Witten,
Superstring Theory, 2 vols., Cambridge
University Press, 1987.}
\REF\CHSW{P. Candelas, G.T. Horowitz, A. Strominger and E. Witten,
\NPB{258}{85}{46}.}
\REF\DHVW{L. Dixon, J.A. Harvey, C. Vafa and E. Witten, \NPB{261}{85}{678};
\NPB{274}{86}{285}.}
\REF\Narain{K.S. Narain, \PLB{169}{86}{41};
W. Lerche, D. L{\H u}st and A.N. Schellekens,
\NPB{287}{87}{477}.}
\REF\FFF{I. Antoniadis, C. Bachas, and C. Kounnas, Nucl.Phys.{\bf B289}
(1987) 87; I. Antoniadis and C. Bachas, Nucl.Phys.{\bf B298} (1988)
586; H. Kawai, D.C. Lewellen, and S.H.-H. Tye, Phys.Rev.Lett.{\bf57} (1986)
1832; Phys.Rev.{\bf D34} (1986) 3794; Nucl.Phys.{\bf B288} (1987) 1;
R. Bluhm, L. Dolan, and P. Goddard, Nucl.Phys.{\bf B309} (1988) 330.}
\REF\REVAMP{I. Antoniadis, J. Ellis, J. Hagelin, and \DVN, \PLB{231}{89}{65}.}
\REF\SUTHREE{B. Greene {\it{el al.}},
Phys.Lett.{\bf B180} (1986) 69;
Nucl.Phys.{\bf B278} (1986) 667;  {\bf B292} (1987) 606;
R. Arnowitt and  P. Nath, Phys.Rev.{\bf D39} (1989) 2006; {\bf D42}
(1990) 2498; Phys.Rev.Lett. {\bf 62} (1989) 222.}
\REF\SSM{L.E. Iba{\~n}ez {\it{et al.}}, Phys.Lett.
{\bf B191}(1987) 282; A. Font {\it{et al.}},
Phys.Lett. {\bf B210} (1988)
101; A. Font {\it{et al.}},
Nucl.Phys. {\bf B331} (1990)
421; D. Bailin, A. Love and S. Thomas,
Phys.Lett.{\bf B194} (1987) 385;
Nucl.Phys.{\bf B298} (1988) 75; J.A. Casas, E.K. Katehou and C. Mu{\~n}oz,
Nucl.Phys.{\bf B317} (1989) 171.}
\REF\FNY{\AEF, D.V. Nanopoulos and K. Yuan, \NPB{335}{90}{437}.}
\REF\EU{\AEF,  \PLB{278}{92}{131}.}
\REF\TOP{\AEF, \PLB{274}{92}{47}.}
\REF\FIQ{A. Font, L.E. Iba{\~n}ez and F. Quevedo,
 Phys.Lett.{\bf B228} (1989) 79.}
\REF\FN{A.E. Faraggi and D.V. Nanopoulos, \MODA{6}{91}{61}.}
\REF\FBV{J. Ellis, J. Lopez and D.V. Nanopoulos, \PLB{252}{90}{53},
G. Leontaris and T. Tamvakis, \PLB{260}{91}{333}.}
\REF\ALR{I. Antoniadis, G. K. Leontaris and J. Rizos, \PLB{245}{90}{161}.}
\REF\ELNa{J. Ellis, J. Lopez and \DVN, \PLB{245}{90}{375}.}
\REF\LG{T. Gannon and C.S. Lam \PRD{41}{90}{492}.}
\REF\KLN{S. Kalara, J. Lopez and D.V. Nanopoulos,
\PLB{245}{91}{421};
\NPB{353}{91}{650}.}
\REF\naturalness{A.E. Faraggi and D.V. Nanopoulos,
Texas A \& M University preprint CTP--TAMU--78, ACT--15;
\AEF, Ph.D thesis, CTP--TAMU--20/91, ACT--31.}
\REF\ACN{N Sakai and T. Yanagida, \NPB{197}{82}{533};
R. Arnowitt, A.H. Chamsdine and P. Nath, \PRD{32}{85}{2348}.}
\REF\YUKAWA{\AEF, WIS--91/83/NOV--PH.}
\REF\DSW{M. Dine, N. Seiberg and E. Witten, Nucl.Phys.{\bf B289} (1987) 585.}
\REF\AEHN{I. Antoniadis, J. Ellis, A.B. Lahanas,
and \DVN, \PLB{241}{90}{24}.}
\REF\NMASSES{\AEF, \PLB{245}{91}{437}.}

\singlespace
\rl{WIS--92/16/FEB--PH}
\rl\today
\normalspace
\medskip
\titlestyle{\bf Construction of Realistic
Standard--like Models in the Free Fermionic
Superstring Formulation}
\author{Alon E. Faraggi{\footnote*{e--mail address: fhalon@weizmann.bitnet}}}
\medskip
\centerline {Department of Physics, Weizmann Institute of Science}
\centerline {Rehovot 76100, Israel}
\titlestyle{ABSTRACT}

I discuss in detail the
construction of realistic superstring standard--like models
in the four dimensional free fermionic formulation.
The analysis results in a restricted class of models with unique
characteristics: (i) Three and only three generations of chiral fermions
with their superpartners and the correct Standard Model quantum numbers.
(ii) Proton decay from dimension four and dimension five operators
is suppressed due to gauged $U(1)$ symmetries.
(iii) There exist Higgs doublets from two distinct sectors, which
can generate realistic symmetry breaking.
(iv) These  models explain the top--bottom mass
hierarchy. At the trilinear level of the superpotential
only the top quark gets a non vanishing mass term. The bottom quark
and the lighter quarks and leptons get their mass terms from
non renormalizable terms. This result is correlated with the
requirement of a supersymmetric vacuum at the Planck scale. (v)
The models predict the existence of small hidden gauge groups,
like $SU(3)$, with matter spectrum in vector representations.

\titlestyle{To appear in Nucl.Phys.B.}

\singlespace
\vskip 0.5cm
\endpage
\normalspace
\pagenumber 1

\centerline{\bf 1. Introduction}

The quest of theoretical physics in recent years has been the unification
of all known fundamental interactions into one, consistent,
theoretical formulation. Although the main prediction of Unified
Theories, proton decay, has not yet been observed, calculations of
$\sin^2\theta_W$ and of the mass ratio $m_b\over{m_\tau}$ support their
validity. Recent calculations  [\EKN] seem to
favor supersymmetric unification
versus non supersymmetric unification. Superstring theory [\GSW]
is a unique candidate for the
consistent unification of gravity with the gauge interactions, but lacks
experimental support for its existence.

Initially it was believed that for
its consistency the superstring had to
be embedded in ten space--time dimensions and then the extra
dimensions had to be compactified on a Calabi-Yau [2] manifold or
on an orbifold [3] . Further
study revealed that one could formulate a consistent string theory directly
in four space--time dimensions by identifying the extra degrees of freedom
as either bosonic [4] or fermionic [5,6] internal degrees of freedom.

On the other hand the Standard--Model  agrees with all experimental
observations to date, but leaves many questions unresolved. Among them
are the fermion mass hierarchy, the number of chiral generations, the
origin of fundamental scales, etc. These problems find natural explanations
in superstring theories.  Therefore, an important task
is to connect the superstring with the Standard--Model.
This task is obscured by the enormous number of
candidate string models and our ignorance of the mechanism which selects
the unique model.
Two approaches can be followed to connect the superstring with
the Standard--Model. One is to use a GUT model with an intermediate
energy scale. Many attempts have been made in this direction
 and most notable are the
flipped $SU(5)$ [\REVAMP] and the $SU(3)^3$ [\SUTHREE] models.
The second possibility is to derive the Standard--Model directly
from the superstring, without any non--abelian gauge symmetry at
an intermediate energy scale [\SSM,\FNY,\EU,\TOP].

In Ref.
[\FNY,\EU,\TOP] preliminary attempts to derive the Standard--Model
directly from superstring theory were made in the free fermionic
formulation.
However, all these attempts consist of
isolated examples and a systematic presentation is still
lacking.  In this paper I try to fill this gap.
Lacking a dynamical mechanism which singles out
the unique string model, it is naive to expect that a particular
example will turn out to be the correct model.
However, by investigating a whole class of models we can extract
the general properties of
 these models and their low energy phenomenological characteristics.
The free fermionic formulation is chosen due to
its unique properties. First, it is
formulated directly in four space--time dimensions. Second, it is an exact
conformal field theory which gives us the advantage of using the
powerful calculational tools of conformal field theory, yielding
highly predictive models. Finally, it is formulated at the
self--dual point in the compactified space which enhances space--time
gauge symmetries from $U(1)$ to $SU(2)$.

 I present
a detailed discussion of the spin structure basis vectors
and the implications on low energy phenomenology.
I impose the following phenomenological constraints  on a
possible superstring standard--like model:

\parindent=-15pt

1. The gauge group is $SU(3)_C\times SU(2)_L\times {U(1)^n}\times$hidden,
with $N=1$ space-time supersymmetry.

2. Three generations of chiral fermions
and their superpartners, with the correct quantum numbers
under
{\hskip .5cm} ${SU(3)_C\times SU(2)_L\times U(1)_Y}$.

3. The spectrum should contain Higgs doublets that can  produce
realistic gauge symmetry breaking.

4. Anomaly cancellation, apart from a single ``anomalous" U(1)
which is  canceled by  application of the
Dine--Seiberg--Witten (DSW) mechanism.

\bigskip
\parindent=15pt

The focus on the standard--like models in particular is motivated by the
following consideration. In the most general supersymmetric standard model
the dimension four operators,
${\eta_1}{u_{L}^C}{d_{L}^C}{d_{L}^C}+
{\eta_2}{d_{L}^C}QL$, mediate rapid proton decay.
Traditionally in supersymmetric models, one imposes R symmetries on the
spectrum to avoid this problem. In the context of superstring theories
these discrete symmetries are usually not found [\FIQ].
These dimension
four operators are forbidden if the gauge symmetry of the
Standard Model is extended by a $U(1)$ symmetry, which
is a combination of, $B-L$, baryon
minus lepton number,  and $T_{3_R}$, and is exactly the
additional $U(1)$ that is derived in the superstring
standard--like models. The dimension four operators may still
appear from the nonrenormalizable terms
$${\eta_1}({u_{L}^C}{d_{L}^C}{d_{L}^C}N_L^C)\Phi+
{\eta_2}({d_{L}^C}QLN_L^C)\Phi,$$
 where $\Phi$ is a combination of fields which
fixes the string selection rules and gets a VEV of $O(m_{pl})$.
The ratio
${\langle{N_L^c}\rangle}\over{M_{pl}}$ controls the rate
of proton decay. While in the standard--like models
this problem can be evaded either by keeping $B-L$ gauged down to low
energies [\FN], or by simply keeping
${\langle{N_L^c}\rangle}=0$,
in superstring models based on an intermediate GUT symmetry,
the problem is more difficult as $N_L^c$ is necessarily
used to  break the GUT symmetry [\FBV].

The paper is organized as follows. In section 2, I review the
basic tools needed for the construction of models in the free fermionic
formulation. In section 3, I emphasize the special role played by
 the first five vectors in the basis
that spans the models. I argue that the important functions of this set
make it a unique set. In sections 4 and 5 I discuss the construction of
standard--like models and their unique characteristics. In section
6, I discuss some of the phenomenology which is expected to arise from these
models.

\bigskip

\centerline{\bf 2. Basic tools for model building}

In the free fermionic formulation of the heterotic string
in four dimensions all the world--sheet
degrees of freedom  required to cancel
the conformal anomaly are represented  in terms of free fermions
propagating on the string world--sheet.
For the left--movers  (world--sheet supersymmetric) one has the
usual space--time fields $X^\mu$, $\psi^\mu$, ($\mu=0,1,2,3$),
and in addition the following eighteen real free fermion fields:
$\chi^I,y^I,\omega^I$  $(I=1,\cdots,6)$, transforming as the adjoint
representation of $SU(2)^6$. The supercurrent
is given in terms of these fields as follows
$$T_F(z) = \psi^\mu\partial_zX_\mu + {\sum_{i=1}^6}\chi^iy^i\omega^i.\eqno(1)$$
For the right movers we have ${\bar X}^\mu$ and 44 real free fermion fields:
${\bar\phi}^a$, $a=1,\cdots,44.$
Under parallel transport around a noncontractible loop the
fermionic states pick up a phase. A model in this construction
is defined by a set of  basis vectors of boundary conditions for all
world--sheet fermions. These basis vectors are constrained by the string
consistency requirements (e.g. modular invariance) and
completely determine the vacuum structure of
the model. The physical spectrum is obtained by applying the generalized
GSO projections. The low energy effective field theory is obtained
by S--matrix elements between external states. The Yukawa couplings
and higher order nonrenormalizable terms in the superpotential
are obtained by calculating corralators between vertex operators.
For a corralator to be nonvanishing all the symmetries of the model must
be conserved. Thus, the boundary condition vectors determine the
phenomenology of the model.

The class of spin structure
models which
I  investigate here are generated by a basis of $Z_2^7\times Z_4$. The basis
generates an additive group
$\Xi={\sum_{k}}n_kb_k$, where
$n_k=0,\dots,N_{Z_k}-1$.
The physical states in the Hilbert space, of a given sector
$\alpha\epsilon\Xi$,
are obtained [\FFF] by acting on the vacuum
$\vert 0\rangle_{\alpha}$ with bosonic, and
fermionic operators with frequencies $\nu_f={{1+\alpha(f)}\over 2}$,
and $\nu_{f^*}={{1-\alpha(f)}\over 2}$, for $f$ and $f^*$, respectively .
The states satisfy the Virasoro condition:
$$M_L^2=-{1\over 2}+{{{\alpha_L}\cdot{\alpha_L}}\over 8}+N_L=-1+
{{{\alpha_R}\cdot{\alpha_R}}\over 8}+N_R=M_R^2\eqno (2)$$
where
$\alpha=(\alpha_L;\alpha_R)$ is a sector in the additive
group  and $N_L={\sum_{f}}({\nu_{_L}}) ;{\hskip .2cm}
N_R={\sum_{f}}({\nu_{_R}})$.

The only states $\vert{s}\rangle_\alpha$
which contribute to the massless spectrum are
those which satisfy the
generalized GSO projections  [\FFF],
$$\left\{e^{i\pi({b_i}F_\alpha)}-
{\delta_\alpha}c^*\left(\matrix{\alpha\cr
                 b_i\cr}\right)\right\}\vert{s}\rangle_\alpha=0\eqno(3a)$$
with $$(b_i{F_\alpha})\equiv\{\sum_{real+complex\atop{left}}-
\sum_{real+complex\atop{right}}\}(b_i(f)F_\alpha(f)),\eqno(3b)$$
where $F_\alpha(f)$ is a fermion number operator counting each mode of
$f$ once (and if $f$ is complex, $f^*$ minus once).
For periodic fermions the vacuum is a spinor in order to represent the Clifford
algebra of the corresponding zero modes. For each periodic complex fermion f,
there are two degenerate vacua ${\vert +\rangle},{\vert -\rangle}$ ,
annihilated by the zero modes $f_0$ and
${{f_0}^*}$ and with fermion numbers  $F(f)=0,-1$, respectively. The $U(1)$
charges, Q(f), with respect to the unbroken Cartan generators of the four
dimensional gauge group, which are in one to one correspondence with the $U(1)$
currents ${f^*}f$ for each complex fermion f, are given by:
 $${Q(f) = {1\over 2}\alpha(f) + F(f)}\eqno (4)$$
where $\alpha(f)$ is the boundary condition of the world--sheet fermion $f$
 in the sector $\alpha$.

To analyze the massless spectrum, I have written a FORTRAN
program.
The program takes as input the basis vectors $B=\{b_1,\cdots,b_8\}$,
 and the GSO coefficients $c\left(\matrix{b_i\cr
                                    b_j\cr}\right)$, $(i,j=1,\cdots,8)$.
The program checks the
modular invariance rules, spans the additive group
$\Xi={\sum_j}{n_jb_j};$ ($j=1,...,8$), selects the
sectors in $\Xi$ which lead to massless states and performs the GSO
projections. It calculates the traces of the $U(1)$ symmetries and evaluates
trilevel and higher order terms in the
superpotential. The program was tested on the existing examples
in the free fermionic formulation  [\REVAMP,\FNY,\ALR].
This unique program enables the exploration of a wider range of models
rather than specific isolated examples. Combined with
the standard techniques for evaluating non vanishing corralators
and renormalization group equations, it provides powerful machinery
for studying the phenomenology of the superstring models.
\bigskip

\centerline{\bf 3. The NAHE set}

The first five vectors (including the vector {\bf 1}) in the basis are
$$\eqalignno{S&=({\underbrace{1,\cdots,1}_{{\psi^\mu},
{\chi^{1..6}}}},0,\cdots,0
\vert 0,\cdots,0).&(5a)\cr
b_1&=({\underbrace{1,\cdots\cdots\cdots,1}_
{{\psi^\mu},{\chi^{12}},y^{3,...,6},{\bar y}^{3,...,6}}},0,\cdots,0\vert
{\underbrace{1,\cdots,1}_{{\bar\psi}^{1,...,5},
{\bar\eta}^1}},0,\cdots,0).&(5b)\cr
b_2&=({\underbrace{1,\cdots\cdots\cdots\cdots\cdots,1}_
{{\psi^\mu},{\chi^{34}},{y^{1,2}},
{\omega^{5,6}},{{\bar y}^{1,2}}{{\bar\omega}^{5,6}}}}
,0,\cdots,0\vert
{\underbrace{1,\cdots,1}_{{{\bar\psi}^{1,...,5}},{\bar\eta}^2}}
,0,\cdots,0).&(5c)\cr
b_3&=({\underbrace{1,\cdots\cdots\cdots\cdots\cdots,1}_
{{\psi^\mu},{\chi^{56}},{\omega^{1,\cdots,4}},
{{\bar\omega}^{1,\cdots,4}}}},0,\cdots,0
\vert {\underbrace{1,\cdots,1}_{{\bar\psi}^{1,...,5},
{\bar\eta}^3}},0,\cdots,0).&(5d)\cr}$$
with the choice of generalized GSO projections
$$c\left(\matrix{b_i\cr
                                    b_j\cr}\right)=
c\left(\matrix{b_i\cr
                                    S\cr}\right)=
c\left(\matrix{1\cr
                                    1\cr}\right)=-1,$$
and the others given by modular invariance.
This set is reffered to as the NAHE{\footnote*{This set was first
constructed by Nanopoulos, Antoniadis, Hagelin and Ellis  (NAHE)
in the construction
of  the flipped $SU(5)$.  {\it nahe}=pretty, in
Hebrew.}} set.
The NAHE set is common to all  the realistic models constructed  in the
free fermionic formulation [\REVAMP,\FNY,\ALR,\EU,\TOP] and is a basic set
common to
all the models which I present.
The sector {\bf S} generates  $N=4$ space--time supersymmetry, which is broken
to $N=2$ and $N=1$ space--time supersymmetry by $b_1$ and $b_2$, respectively.
Restricting $b_j\cdot S=0$mod$2$, and $c\left(\matrix{S\cr
                                                b_j\cr}\right)=\delta_{b_j}$,
for all basis vector $b_j\epsilon{B}$ guarantees the existence of
$N=1$ space--time supersymmetry. The superpartners from a given sector
$\alpha\epsilon\Xi$ are obtained from the sector $S+\alpha$.
The gauge group after the NAHE set is $SO(10)\times
E_8\times SO(6)^3$ with $N=1$ space--time supersymmetry.
The three $SO(6)$
symmetries are horizontal, generational dependent, symmetries.
The vectors $b_1$, $b_2$ and $b_3$ of the NAHE set perform several
functions:

\parindent=-15pt

1. They produce the chiral generations.

2. They perform a ``chirality operation".
 To obtain from a  given sector $b_j$ a full spinorial 16 of $SO(10)$
with the same chirality, a second vector is needed in the basis.
$\psi^\mu,{\bar\psi}^{1,\cdots,5}$ are periodic in both vectors
and the intersection between the remaining boundary conditions is empty.

3. They separate the hidden sector from the observable sector.

\parindent=15pt

At the level of the NAHE set, each
sector $b_1$, $b_2$ and  $b_3$  give rise to 16 spinorial 16 of $SO(10)$.
The internal $44$ right--moving  fermionic states
are divided in the following way:
${\bar\psi}^{1,\cdots,5}$ are complex and produce the observable $SO(10)$
symmetry;
${\bar\phi}^{1,\cdots,8}$ are complex and produce the hidden $E_8$ gauge
group;
$\{{\bar\eta}^1,{\bar y}^{3,\cdots,6}\}$, $\{{\bar\eta}^2,{\bar y}^{1,2}
,{\bar\omega}^{5,6}\}$, $\{{\bar\eta}^3,{\bar\omega}^{1,\cdots,4}\}$
 give rise to the three horizontal $SO(6)$ symmetries.
The left--moving $\{y,\omega\}$ states
are divided to,
$\{{y}^{3,\cdots,6}\}$, $\{{y}^{1,2}
,{\omega}^{5,6}\}$, $\{{\omega}^{1,\cdots,4}\}$.
The left--moving $\chi^{12},\chi^{34},\chi^{56}$ states
carry the supersymmetry charges.

\parindent=15pt

 The
Neveu--Schwarz sector produces the massless scalar states
$$\eqalignno{&{\chi_{1\over2}^{12}}{{\bar\psi}_{1\over2}^{{1\cdots5}}}
\{{\bar\eta}^1_{1\over2},{\bar y}^{3\cdots6}_{1\over2}\},{\hskip 1in}
\chi^{12}_{1\over2}\{{\bar\eta}^3_{1\over2}{\bar\omega}^{1\cdots4}_{1\over2}
\}\{{\bar\eta}^2_{1\over2},{\bar y}^{1,2}_{1\over2},
{\bar\omega}_{1\over2}^{5,6}\}&(6a,b)\cr
&\chi^{34}{{\bar\psi}^{1\cdots5}_{1\over2}}\{{\bar\eta}^2,
{\bar y}^{1,2}_{1\over2}
{\bar\omega}^{5,6}_{1\over2}\},{\hskip 2cm}
\chi^{34}_{1\over2}\{{\bar\eta}^1_{1\over2},
{\bar y}^{3\cdots6}_{1\over2}\}
\{{\bar\eta}^3_{1\over2},{\bar\omega}^{1\cdots4}_{1\over2}\}&(6c,d)\cr
&\chi^{56}_{1\over2}{{\bar\psi}^{1\cdots5}_{1\over2}}
\{{\bar\eta}^3_{1\over2},{\bar\omega}^{1\cdots4}_{1\over2}\},{\hskip 1in}
\chi^{56}_{1\over2}\{{\bar\eta}^1_{1\over2},{\bar y}^{3,\cdots,6}_{1\over2}\}
\{{\bar\eta}^2_{1\over2}{\bar y}^{1,2}_{1\over2}
{\bar\omega}_{1\over2}^{5,6}\}&(6e,f)}$$
(and their complex conjugates).
The states from the first column give rise to $5$ and ${\bar 5}$ under
$SU(5)$. These produce Higgs doublets which are used  to break
the electroweak symmetry. The states in the second column  are singlets under
the $SO(10)$ symmetry and are used in the application of the DSW mechanism.

Another consequence of the NAHE set is observed by extending the $SO(10)$
symmetry to $E_6$. Adding the vector
$$X=(0,\cdots,0\vert{\underbrace{1,\cdots,1}_{{\psi^{1,\cdots,5}},
{\eta^{1,2,3}}}},0,\cdots,0)\eqno(7)$$
to the NAHE set, extends the gauge symmetry to
$E_6\times U(1)^2\times SO(4)^3$.
The sectors $(b_1;b_1+X)$, $(b_2;b_2+X)$ and $(b_3;b_3+X)$
each give eight $27$ of $E_6$. The $(NS;NS+X)$ sector gives in
addition to the vector bosons and spin two states, three copies of
scalar representations in $27+{\bar {27}}$ of $E_6$.

In this model the only internal fermionic states which count the
multiplets of $E_6$ are the real internal fermions $\{y,w\vert{\bar y},
{\bar\omega}\}$. This is observed by writing the degenerate vacuum
of the sectors $b_j$ in a combinatorial notation. The vacuum of the sectors
$b_j$  contains twelve periodic fermions. Each periodic fermion
gives rise to a two dimensional degenerate vacuum $\vert{+}\rangle$ and
$\vert{-}\rangle$ with fermion numbers $0$ and $-1$, respectively.
The GSO operator, Eq. (3), is a generalized parity, operator which
selects states with definite parity. From Eq. (3) and after applying the
GSO projections, we can write the degenerate vacuum of the sector
$b_1$ in combinatorial form
$$\eqalignno{\left[\left(\matrix{4\cr
                                    0\cr}\right)+
\left(\matrix{4\cr
                                    2\cr}\right)+
\left(\matrix{4\cr
                                    4\cr}\right)\right]
\left\{\left(\matrix{2\cr
                                    0\cr}\right)\right.
&\left[\left(\matrix{5\cr
                                    0\cr}\right)+
\left(\matrix{5\cr
                                    2\cr}\right)+
\left(\matrix{5\cr
                                    4\cr}\right)\right]
\left(\matrix{1\cr
                                    0\cr}\right)\cr
+\left(\matrix{2\cr
                                    2\cr}\right)
&\left[\left(\matrix{5\cr
                                    1\cr}\right)+
\left(\matrix{5\cr
                                    3\cr}\right)+
\left(\matrix{5\cr
                                    5\cr}\right)\right]\left.
\left(\matrix{1\cr
                                    1\cr}\right)\right\}&(8)\cr}$$
where
$4=\{y^3y^4,y^5y^6,{\bar y}^3{\bar y}^4,
{\bar y}^5{\bar y}^6\}$, $2=\{\psi^\mu,\chi^{12}\}$,
$5=\{{\bar\psi}^{1,\cdots,5}\}$ and $1=\{{\bar\eta}^1\}$.
The combinatorial factor counts the number of $\vert{-}\rangle$ in
a given state. The two terms in the curly brackets correspond to the two
components of a Weyl spinor.  The $10+1$ in the $27$ of $E_6$ are
obtained from the sector $b_j+X$.
{}From Eq. (8) it is observed that the states
which count the multiplicities of $E_6$ are the internal
fermionic states $\{y^{3,\cdots,6}\vert{\bar y}^{3,\cdots,6}\}$.
A similar result is
obtained for the sectors $b_2$ and $b_3$ with $\{y^{1,2},\omega^{5,6}
\vert{\bar y}^{1,2},{\bar\omega}^{5,6}\}$
and $\{\omega^{1,\cdots,4}\vert{\bar\omega}^{1,\cdots,4}\}$
respectively, which suggests that
these twelve states correspond to a six dimensional
compactified orbifold with Euler characteristic equal to 48.

I would like to emphasize that the functions 1 and 2 above make the
partial set $\{{\bf 1},S,b_1,b_2\}$ of the NAHE set  a
completely general set. Indeed,
this partial set is common, in one form or another, to all the constructions
in the free fermionic formulation. The minimal way to obtain a well
defined hidden gauge group [\LG] is by adding the vector $b_3$ to this set,
which makes the NAHE set a unique set. The analysis of models beyond the NAHE
set is reduced, almost entirely, to the study of the boundary conditions
of the  real fermions $\{y^{1,\cdots,6},\omega^{1,\cdots,6}\vert
{\bar y}^{1,\cdots,6},{\bar\omega}^{1,\cdots,6}\}$, and is simplified
considerably. In the language of conformal field theory these real fermions
correspond to the left right symmetric internal conformal field theory.
As I will show bellow many of the phenomenological
implications are determined by the boundary conditions of these real
fermions.

\bigskip
\centerline{\bf 4.  Beyond the NAHE set}

In the following I employ  a table notation which
emphasizes the division of the internal fermionic
states according to their division by the NAHE set.
The set of real fermions $\{y,w\vert{\bar y},{\bar\omega}\}$
plays an important role in the low energy properties of the
standard--like models. In the table,
the real fermionic states $\{y,w\vert{\bar y},{\bar\omega}\}$
are divided according to their division by the NAHE set.
The pairing of real fermions into complex fermions or into
Ising model sigma operators is noted in the table.
The entries in the table represent the boundary conditions
in a basis vector for all the fermionic states.
The basis vectors in a given table are the three
basis vectors which extend the NAHE
set.

\noindent{\bf 4.1  The observable gauge group and its symmetries}

The properties of the observable sector
are determined by the set of internal fermionic
states, $\{\psi^\mu,\chi^i,y^i,\omega^i\vert {\bar y}^i,{\bar\omega}^i,
{\bar\psi}^{1\cdots5},{\bar\eta}^1,{\bar\eta}^2,{\bar\eta}^3\}$,
$(i=1,\cdots6)$. Different
models differ, with respect to the properties of the  observable
sector, by the assignment of boundary conditions in the basis vectors
which extend the NAHE set. A strong constraint on the possible
gauge group comes from the absence of adjoint representations
in the
massless spectrum of level one Kac--Moody algebra [\ELNa].
Therefore
the $SO(10)$ symmetry
has to be broken to one of its subgroups $SU(5)\times U(1)$,
$SO(6)\times SO(4)$ or
$SU(3)\times SU(2)\times U(1)_{B-L}\times U(1)_{T_{3_R}}$.
This is achieved by the assignment of boundary conditions to the set
${\bar\psi}^{1\cdots5}_{1\over2}$:

1. $b\{{{\bar\psi}_{1\over2}^{1\cdots5}}\}=
\{{1\over2}{1\over2}{1\over2}{1\over2}
{1\over2}\}\Rightarrow SU(5)\times U(1)$,

2. $b\{{{\bar\psi}_{1\over2}^{1\cdots5}}\}=\{1 1 1 0 0\}
  \Rightarrow SO(6)\times SO(4)$.

To
  break the $SO(10)$ symmetry to
$SU(3)\times SU(2)\times
U(1)_C\times U(1)_L$\footnote*{$U(1)_C={3\over2}U(1)_{B-L};
U(1)_L=2U(1)_{T_{3_R}}.$}
both steps, 1 and 2, are used, in two separate basis vectors
\footnote\dag{There is a second possibility
of using two $Z_4$ twists with $b\{{{\bar\psi}^{1\cdots5}}\}=
\{{1\over2}{1\over2}{1\over2}{1\over2}
{1\over2}\}$ and $b^{\prime}\{{{\bar\psi}^{1\cdots5}}\}$
$ =\{{1\over2}{1\over2}{1\over2}-{1\over2}
-{1\over2}\}$, however it is found that the massless spectrum in the two cases
is equivalent.}. The $SO(10)$ symmetry has to be broken in at least two of the
three vectors which extend the NAHE set. Models in which the $SO(10)$ symmetry
is broken in all three vectors are  possible.

The weak hypercharge is given by the combination
$U(1)_Y={1\over3}U(1)_C+{1\over2}U(1)_L$. The orthogonal combination
is given by $U(1)_{Z^\prime}=U(1)_C-U(1)_L$.

The number of horizontal $U(1)$ symmetries depends on the assignment of
boundary conditions and differs between models. All models have at
least three horizontal $U(1)$ symmetries, denoted by $U(1)_{r_j}$
$(j=1,2,3)$,  which correspond to the
right--moving world--sheet
currents ${\bar\eta}^1_{1\over2}{\bar\eta}^{1^*}_{1\over2}$,
${\bar\eta}^2_{1\over2}{\bar\eta}^{2^*}_{1\over2}$ and
${\bar\eta}^3_{1\over2}{\bar\eta}^{3^*}_{1\over2}$.
These complex fermionic states
are twisted by a $Z_4$ twist. This twist is necessary to keep the two
Weyl spinor components of the chiral fermions in the spectrum.
Additional horizontal $U(1)$ symmetries, denoted by $U(1)_{r_j}$
$(j=4,5,...)$,
arise by pairing two real fermions from the sets
 $\{{\bar y}^{3,\cdots,6}\}$, $\{{\bar y}^{1,2},{\bar\omega}^{5,6}\}$ and
$\{{\bar\omega}^{1,\cdots,4}\}$. The final observable gauge group depends on
the number of such pairings.

For each of these complexified right--moving fermions correspond a
left--moving complexified fermion from the sets
$\{y^{3\cdots6}\}$, $\{y^{1,2},\omega^{5,6}\}$ and
$\{\omega^{1\cdots4}\}$. The complexification of left--moving fermions,
and the assignment of boundary conditions to them, is further constrained
by the world--sheet supercurrent, Eq. (1). In the fermionic formulation and
with the supersymmetry generator of the NAHE set, the boundary conditions
of any  ($\chi^I,y^I,\omega^I$) triplet can belong only to
$\{(1,1,0);(1,0,1);(0,1,1);(0,0,0)\}$ for space--time bosons
and to  $\{(1,0,0);(0,1,0)$ $;(0,0,1);(1,1,1)\}$
for space--time fermions.
Each complexified left--moving
fermion gives rise to a global $U(1)$ symmetry, denoted by
$U(1)_{\ell_j}$
 $(j=4,5,...)$.
As will be shown bellow these additional
horizontal symmetries play an important role in the phenomenology of the
massless  spectrum.
Three additional left--moving global $U(1)$ symmetries, denoted by
$U(1)_{\ell_j}$ $(j=1,2,3)$,  arise from the
charges of the supersymmetry generator: $\chi^{12}$, $\chi^{34}$
and $\chi^{56}$.

If all right--moving (and hence all left--moving) fermions were complex,
the gauge group would have rank 22. The rank is reduced by pairing a
left--moving fermion $(f)$ with a real right--moving fermion ${\bar f}$
to form an Ising model sigma operator. These are denoted by
$\sigma^i_\pm$ and $\sigma^{\bar i}_\pm$ for ${(y^i{\bar y}^i)}_\pm$
and ${(\omega^i{\bar\omega}^i)}_\pm$, respectively.
For a corralator between vertex operators
to be non vanishing, the real fermions must produce non zero
Ising model corralators. The symmetries of the Ising model corralators
and of the left moving charges must be checked after all picture
changing have been done. The rules for obtaining the non vanishing
corralators are given in Ref. [\KLN].

\noindent{\bf 4.2  The number of generations}

The question of the number of generations is discussed in detail
in Ref. [\naturalness].
It is argued that the NAHE set  leads to  three
generations as the most natural number of generations.
After the NAHE set, each
sector $b_1$, $b_2$ and $b_3$ give rise to sixteen chiral generations.
The number of generations is determined by the set of real
 fermions $\{y,\omega\vert{\bar y},{\bar\omega}\}$
(the vertical line separates left from right movers).
The reduction to three generations is illustrated in
the model of table 1.

At the level of the NAHE set we have 48 generations.
One half of the generations is projected because of the $1\over2$ twist.
In the $Z_4$ projection
one half of the generations have $\pm1$ signature while the other half
has $\pm{i}$. Therefore one half of the generations is projected by the
$Z_4$ twist. Each of the vectors in table 1 acts nontrivially on the
real part of the
sectors $b_1$, $b_2$ and $b_3$ and reduces the combinatorial factor
of Eq. (8) by a half. Thus, we obtain one generation from each sector
$b_1$, $b_2$ and $b_3$.

It is important to note that if the final gauge group is
$SU(5)\times U(1)$ or $SO(6)\times SO(4)$ two of the additional sectors give
rise to $16+{\bar{16}}$ of $SO(10)$. The net chirality of three generations
is not spoiled.
 If the $SO(10)$  symmetry is broken to  $SU(3)\times SU(2)\times
U(1)^2$, constructions with exactly three generations and no mirror
generations are obtained.

In the notation of table 1, all
the real fermions are paired to form Ising model
operators.  The states ${\bar\eta}^{1,2,3}$ are complex
and are separated from the real fermions by the $1\over2$ twist in the sector
$\gamma$. At least three additional vectors are needed
to  break all the horizontal symmetries which arise from the part of the
real fermions and at the same time reduce the number of generations
to one generation from
each of the sectors $b_1$, $b_2$ and $b_3$. Thus,
the minimal additive group is
$\Xi=Z_2^7\otimes Z_4$. In the model of table 1 all the
real horizontal symmetries
are completely broken and the rank of the final gauge group is 16.
In the models
which I introduce below, some of the real fermions are paired to form
complex fermions and therefore  give rise to additional horizontal $U(1)$
symmetries.

In the free fermionic models the chiral generations, from the sectors
$b_1$, $b_2$ and $b_3$, carry charges under the three horizontal
$U(1)_j$  $(j=1,\cdots,3)$ symmetries. The sign is determined by the product,
$\gamma\cdot{b_j}=odd/even\Rightarrow{U(1)_j=-{1\over2}/{1\over2}}$,
respectively.
In addition to these symmetries the chiral generations
carry charges under the additional horizontal  $U(1)_{r_j}$
$(j=4,5,6)$ symmetries.
For example in the model of table 2 we obtain three chiral generations
$G_j=e_{L_j}^c+u_{L_j}^c+N_{L_j}^c+d_{L_j}^c+
Q_j+L_j$  $(j=1,\cdots,3)$
with the following charges.
{}From the sector $b_1$ we obtain
$${\hskip .2cm}  ({e_L^c}+{u_L^c})_{-{1\over2},0,0,-{1\over2},0,0}+
({d_L^c}+{N_L^c})_{-{1\over2},0,0,{-{1\over2}},0,0}+
(L)_{-{1\over2},0,0,{1\over2},0,0}+(Q)_{-{1\over2},0,0,{1\over2},0,0},
\eqno(9a)$$ from the sector $b_2$
$${\hskip .2cm} ({e_L^c}+{u_L^c})_{0,-{1\over2},0,0,{1\over2},0}+
({N_L^c}+{d_L^c})_{0,-{1\over2},0,0,-{1\over2},0}+
(L)_{0,-{1\over2},0,0,-{1\over2},0}+
(Q)_{0,-{1\over2},0,0,{1\over2},0},\eqno(9b)$$ and from the sector $b_3$
$$ {\hskip .2cm} ({e_L^c}+{u_L^c})_{0,0,{1\over2},0,0,{1\over2}}+
({N_L^c}+{d_L^c})_{0,0,{1\over2},0,0,-{1\over2}}+
(L)_{0,0,{1\over2},0,0,-{1\over2}}+(Q)_{0,0,{1\over2},0,0,{1\over2}}.
\eqno(9c)$$
Where
$$\eqalignno{{e_L^c}&\equiv [(1,{3\over2});(1,1)];{\hskip .6cm}
{u_L^c}\equiv [({\bar 3},-{1\over2});(1,-1)];{\hskip .2cm}
Q\equiv [(3,{1\over2});(2,0)]{\hskip 2cm}
&(10a,b,c)\cr
{N_L^c}&\equiv [(1,{3\over2});(1,-1)];{\hskip .2cm}
{d_L^c}\equiv [({\bar 3},-{1\over2});(1,1)];{\hskip .6cm}
L\equiv [(1,-{3\over2});(2,0)]{\hskip 2cm}
&(10d,e,f)\cr}$$
of $SU(3)_C\times U(1)_C\times SU(2)_L\times U(1)_L$.
The  charges under the
six horizontal $U(1)$ are given in Eqs. (9). Three generations of
chiral fermions are common to all the models which I present.

\bigskip
\noindent{\bf 4.3  Higgs doublets}

The massless spectrum must contain Higgs doublets to give masses to the quarks
and leptons. The Higgs doublets in the free fermionic models
are obtained from
two types of sectors.
The first type are scalar doublets from the Neveu--Schwarz sector which arise
from the scalar representations Eqs. (6). The presence of this scalar doublets
in the massless spectrum is correlated with the additional
$U(1)_{r_j}$ horizontal
symmetries which arise from pairing real right--moving fermions. This pairing
guarantees that both the chiral family from a sector $b_j$ $(j=1,2,3)$, as
well as the corresponding Higgs doublets, $h_j$ and $\bar h_j$,
remain in the massless
spectrum. Otherwise an exclusion principle is observed in the application
of the GSO projection, $\alpha$, which breaks the  $SO(10)$ symmetry to
$SO(6)\times SO(4)$.
If $\alpha\cdot{b_j}=0mod2$  $(j=1,2,3)$, the family from
$b_j$ is in the spectrum and the Higgs  doublet
${\chi_{1\over2}}{{\bar\psi}^{45}_{1\over2}}{{\bar\eta}_{1\over2}^j}
{{\vert}
0\rangle}_0$ is projected out. If $\alpha\cdot{b_j}=1mod2$,
$(j=1,2,3)$, the
family from ${b_j}$ is projected out and the Higgs doublet
${\chi_{1\over2}}{{\bar\psi}^{45}_{1\over2}}
{{\bar\eta}_{1\over2}^j}{{\vert}0\rangle}_0$ is in the spectrum. For every
right--moving $U(1)_{r_j}$ correspond a left--moving global $U(1)_{\ell_j}$
symmetry. The product $\alpha\cdot{b_j}=0mod2$ imposes the constraint
$\vert\alpha(U(1)_{\ell_j})-\alpha(U(1)_{r_j})\vert=1$, which insures
that both the chiral generations $G_j$ and the Higgs doublets
$h_j$, $\bar h_j$ remain in the massless spectrum.

To illustrate this dependence I consider the models in tables  1, 2
and 5.
In model  2, the three horizontal ($U(1)_\ell;U(1)_r$)
symmetries, which correspond  to the
world-sheet currents $(y^3y^6;{\bar y}^3{\bar y}^6)$,
$(y^1\omega^6;{\bar y}^1{\bar\omega}^6)$
and $(\omega^1\omega^3;{\bar\omega}^1{\bar\omega}^3)$,
guarantee that the Higgs doublets $h_1$, ${\bar h}_1$, $h_2$, ${\bar h}_2$
and $h_3$, ${\bar h}_3$, as well as the
chiral generations from the sectors $b_1$, $b_2$ and $b_3$,
remain in the massless spectrum. A similar result is obtained in
models 3 and 4. In model 1 all the real fermions are paired to form
Ising model operators and there are no additional $U(1)$ symmetries
beyond $U(1)_j$ $(j=1,2,3)$. All the Higgs doublets from the
Neveu--Schwarz sector are projected out. In this case the Higgs
triplets $D_1$, $\bar D_1$, $D_2$, $\bar D_2$ and $D_3$, $\bar D_3$
from Eqs. (6a,c,e) remain in the massless spectrum.
In model 5 we have only one additional horizontal
($U(1)_\ell;U(1)_r$) symmetry which corresponds to the world--sheet currents
$(\omega^2\omega^3;{\bar\omega^2}{\bar\omega^3})$.
Therefore in this model
only one pair of Higgs doublets from the Neveu--Schwarz sector,
$h_3$, ${\bar h}_3$, remains in
the massless spectrum after the GSO projections. In this case we obtain from
Eqs. (6a,c) the Higgs triplets $D_1$, $\bar D_1$ and $D_2$, $\bar D_2$.
Thus, the extra horizontal $U(1)_{r_j}$ symmetries
perform an additional function.
They eliminate  the dangerous Higgs
triplets, $D$ and ${\bar D}$,
which mediate proton decay through  dimension five operators [\ACN].

The horizontal $U(1)_{\ell,r}$ symmetries also
guarantee that the $SU(5)$ singlets
from Eqs. (6b,d,f) remain in the massless spectrum. Thus, in  models
2,3 and 4 we obtain three pairs of singlets $\Phi_{12}$, $\bar\Phi_{12}$,
$\Phi_{34}$, $\bar\Phi_{34}$ and $\Phi_{56}$, $\bar\Phi_{56}$,
while in model 5 we obtain only one pair of singlets,
$\Phi_{12}$, $\bar\Phi_{12}$.

The second type of Higgs doublets is obtained
 from a combination of the basis vectors $\alpha$ and $\beta$
 with some combination of $b_1$, $b_2$ and $b_3$.
 For example in models 3 and 4 they arise from the combination
 $\zeta=b_1+b_2+\alpha+\beta$. In this vector,
$\zeta_R\cdot{\zeta_R}=\zeta_L\cdot{\zeta_L}=4$.
 Therefore the massless states are obtained by acting on the vacuum
 with one right--moving fermionic oscillator. The states in this sector
 transform only under the observable gauge group. The presence of these
 states in the massless spectrum, and consequently of this vector combination
 in the additive group, is essential for the application of the DSW mechanism
 and for obtaining realistic phenomenology.
 Requiring the existence of this combination in the additive group
 imposes an additional strong constraint on the allowed
 basis vectors.
 For example, in the model of Ref. [\FNY], it is impossible to obtain
 such a combination. The reason is the specific pairing of the left--moving
 real fermions, $y^3y^6$, $y^1\omega^6$ and $\omega^1\omega^3$.
These
 pairings guarantee that both the chiral fermions from the sectors
  $b_j$ as well as the corresponding  Higgs doublets $h_j$, $\bar h_j$
  are in the massless spectrum. However, the restrictions on the
  boundary conditions of the left--moving triplets
$(\chi^I,y^I,\omega^I)$,
  forbid the construction of a combination like $\zeta$. Therefore in all
  the models with this pairing of left--moving fermions, these type of
  doublets and singlets  does not exist. In models 3 and 4 the pairing of
  left--moving fermions is $y^3y^6$, $y^1\omega^5$ and $\omega^2\omega^4$.
  In this case a vector of the form of $\zeta$ is obtained.
The  singlets and doublets from this sector
play an important role in the application of the
DSW mechanism and in the generation of the fermion
mass hierarchy.

\bigskip
\noindent{\bf 4.4  Yukawa couplings}

The determination of trilevel Yukawa couplings, for the chiral generations
from the sectors $b_1, b_2$ and $b_3$,  depends on the assignment of
boundary conditions for the set of fermions $\{{y^{1,\cdots,6},
{\omega}^{1,\cdots,6}|{\bar y}^{1,\cdots6},
{\bar\omega}^{1,\cdots,6}}\}$. To illustrate this
dependence I consider the model of table 2.
The full massless spectrum of this model is presented in Ref. [\FNY].
The sectors $b_1, b_2$ and $b_3$ give rise to three chiral generations.
{}From the Neveu-Schwarz sector, three pairs of $SU(2)_L$
scalar doublets are obtained.

The basis of table 2 leads to the following trilevel mass terms for the
states from the sectors $b_1, b_2$ and $b_3$:
$$\{({u_{L_1}^c}Q_1{\bar h}_1+{N_{L_1}^c}L_1{\bar h}_1+
{d_{L_2}^c}Q_2h_2+{e_{L_2}^c}L_2h_2+
{e_{L_3}^c}L_3h_3+{d_{L_3}^c}Q_3h_3).\eqno(11)$$

The non vanishing Yukawa couplings for the
states from a given sector $b_1, b_2$
or $b_3$ depend on the assignment of boundary conditions for
the real  fermions
in the vector $\gamma$. For example,
for the sector $b_1$,
${{\bar y}^3}{{\bar y}^6}$ receives  periodic boundary conditions,
${\gamma}({\bar y}^3{\bar y}^6)=1$ while ${y^3}{y^6}$ receives antiperiodic
boundary conditions ${\gamma}({y^3}{y^6})=0$.
This asymmetry leads to a non
vanishing Yukawa coupling for the $+{2\over3}$
charged quark and for the neutral lepton
from the sector $b_1$.
On the other hand, examination of the real fermion
states from the sectors $b_2$ and $b_3$ reveals that for both sectors the
corresponding charges are symmetric in the vector $\gamma$.
${\gamma}({y^1}{\omega^6})={\gamma}({\bar y}^1{\bar\omega}^6)=1$ and
${\gamma}({\omega^1}{\omega^3})={\beta}({\bar\omega}^1{\bar\omega}^3)=0$.
This symmetry leads  to a non
vanishing trilevel Yukawa coupling for the
$-{1\over 3}$ charged quark and for the charged lepton.
In Ref. [\YUKAWA], I prove that in the symmetric case,
$\vert\gamma(U(1)_{\ell_{j+3}})-\gamma(U(1)_{r_{j+3}})\vert=0$,
trilevel mass terms are
possible only for $-{1\over3}$
type quarks  while in the asymmetric case,
$\vert\gamma(U(1)_{\ell_{j+3}})-\gamma(U(1)_{r_{j+3}})\vert=1$,
trilevel mass terms are possible only for $+{2\over3}$
type quarks.
The proof is based on showing that, for the states
from a sector $b_j$, in the symmetric case only $-{1\over3}$
type quarks form trilevel mass terms which are invariant under $U(1)_j$,
while in  the asymmetric case only $+{2\over3}$
type quarks form trilevel mass terms which are invariant under $U(1)_j$,
$(j=1,2,3)$.

{}From this result it follows that, depending on the assignment of boundary
conditions in the vector $\gamma$, it is possible to construct models with
trilevel Yukawa couplings for $+{2\over3}$ charged quarks as well as for
$-{1\over3}$ charged quarks and  for charged leptons.  Apriori,
the Yukawa couplings for all the heaviest generation states
 can be  obtained from trilevel
terms in the superpotential. I will refer to this type of models as
type {\bf I}
models. On the other hand, it is possible to construct models in which only
one type of Yukawa coupling is obtained at trilevel.
For example, in  models 3 and 4,
only $+{2\over3}$ charged quarks get
a non vanishing trilevel Yukawa coupling.  I will refer to this
models as type {\bf II} models. In the next section I argue that the
requirement of a supersymmetric vacuum at the Planck scale may indicate
that only type {\bf II} models are allowed.

I now turn to discuss Yukawa couplings from nonrenormalizable terms in these
models. A realistic string model must produce mass terms for the lighter
quarks and leptons. The next step is to identify the mass terms for the
bottom quark and for the tau lepton. In type {\bf I} models  these
mass terms arise from trilevel terms. In type {\bf II}
models these terms may arise
from quartic, quintic or higher order terms. As it determines the non
vanishing trilevel Yukawa couplings, the set of real fermions
$\{y,w\vert{\bar y},{\bar\omega}\}$ determines the non vanishing
mass terms from higher orders.

The rules for obtaining the non vanishing higher order terms are given in
Ref. [\KLN].
A non vanishing F term in the superpotential must obey all  the string
selection rules. It must be invariant under all the gauge and global
symmetries. In addition the real fermions must produce non zero
Ising model corralators for a non renormalizable term to be non vanishing.
The symmetries of the Ising model corralators and of the left--moving
global symmetries must be checked after all picture changing have been
done [\KLN].

Examination of the quartic level terms in the model of table 2
reveals that there are no quartic terms which can give rise to
bottom quark and tau lepton mass terms. On the other hand the model of
table 3 does give rise to non vanishing quartic level mass terms for the
bottom quark and for the tau lepton.
These quartic order terms are of the form [\TOP],
$$W_4=\{{d_{L_1}^c}Q_1h_{45}^\prime\Phi_1+{e_{L_1}^c}L_1h_{45}^\prime\Phi_1+
{d_{L_2}^c}Q_2h_{45}^\prime{\bar\Phi}_2
+{e_{L_2}^c}L_2h_{45}^\prime{\bar\Phi}_2\}.\eqno(12)$$
In model 3 nonvanishing mass terms for the bottom quark
and for the tau lepton may be obtained from the following
non vanishing quintic terms,
$$W_5=\{{d_{L_1}^c}Q_1h_{45}\Phi_1^-\xi_2
+{e_{L_1}^c}L_1h_{45}\Phi_1^+\xi_2+
{d_{L_2}^c}Q_2h_{45}{\Phi}_2^-\xi_1
+{e_{L_2}^c}L_2h_{45}{\bar\Phi}_2^-\xi_1\}.\eqno(13)$$
The second type of  Higgs doublets, from the vector combination of
$\alpha+\beta$ plus a combination of $b_1$, $b_2$ and $b_3$,
generate the fermion mass hierarchy in the heaviest generation.
They couple to the bottom quark and to the tau lepton
to form effective Yukawa couplings from the nonrenormalizable terms.
In the application of the DSW mechanism the singlets in Eqs. (12,13) acquire
a VEV. Thus, the
effective Yukawa couplings are suppressed by a
factor of  $(VEV)^n\over{M_{pl}^n}$ relative
to the trilevel terms [\TOP].

\bigskip
\noindent{\bf 4.5  Anomalous $U(1)$}

The massless spectrum of the free fermionic models contains anomaly free and
anomalous $U(1)$ symmetries. The boundary condition vectors and the choice
of GSO phases determine
the anomaly free and anomalous $U(1)$ symmetries.
For example in model 2 the following $U(1)$s are anomalous:
Tr${U_1=-24}$, Tr${U_2=-30}$, Tr${U_3=18}$,
Tr${U_5=6}$, Tr${U_6=6}$ and  Tr${U_8=12}$.
Changing $c\left(\matrix{b_4\cr
                  1\cr}\right)=+1$ to
            $c\left(\matrix{b_4\cr
                  1\cr}\right)=-1,$ changes
the anomalous $U(1)$s to: Tr$U_C=-18$, Tr$U_L=12$,
Tr$U_1=-18$, Tr$U_2=-24$, Tr$U_3=24$,
Tr$U_4=-12$, Tr$U_5=6$, Tr$U_6=6$, Tr$U_7=-6$, Tr$U_8=12$
and Tr$U_9=18$.

The anomalous $U(1)$ is broken by the Dine-Seiberg-Witten mechanism, [\DSW]
in which a potentially large Fayet-Iliopoulos D term  is generated  by
the VEV of the dilaton field $(\phi_D)$.
Such a D term will in general break supersymmetry and destabilize the string
vacuum, unless there is a direction in  the scalar potential
$\phi=\sum_i {{\alpha_i}{\phi_i}}$, which is F flat and also D flat with
respect to the  nonanomalous gauge symmetries and in which $\sum_i {{{Q_i}^A}
{{\vert}{{\alpha_i}}{\vert}}^2}<0$. If such a direction exists, it will
acquire a VEV, canceling the anomalous D term, restoring supersymmetry and
stabilizing the vacuum.
Since the fields corresponding to such a flat direction typically also
carry charges for the non anomalous D terms, a non trivial set of constraints
on the possible choices of VEVs is imposed
and will in general break all of these
symmetries spontaneously.

The set of constraints is summarized in the following set of equations:
$$\eqalignno{{D_A}&={\sum _{k}}{Q_k^A}{\vert\chi_k\vert}^2=
{-g^2e^{\phi_D}\over{192\pi^2}}{Tr(Q_A)}&(14a)\cr
D_j'&=\sum_{k}{Q'}_k^j\vert\chi_k\vert^2=0{\hskip .3cm}
 j=1\cdots5&(14b)\cr
D_j&=\sum_{k}Q_k^j\vert\chi_k\vert^2=0{\hskip .3cm} j=C,L,7,8&(14c)\cr
W&={{\partial W}\over{\partial{\eta_i}}}=0&(14d)\cr}$$
where   ${\chi}_k$ are the fields that get a VEV and ${Q_k}^j$ is their charge
under the $U(1)_j$ symmetry.
The set $\{\eta_j\}$ is the set of fields with
vanishing VEV.
The solution to the set of Eqs.(14) must be positive
definite since ${\vert\chi_k\vert}^2\ge0$.

The set of Eqs. (14) is a non trivial constraint on the allowed models.
To illustrate the difficulty in finding solutions to the set of constraints
I consider the model of table 5.

The observable gauge group of the model is
$SU(3)_C\times U(1)_C\times SU(2)_L\times U(1)_L\times U(1)^4$
and the hidden gauge group is
  $SU(5)_H\times SU(3)_H\times U(1)^2$.
The horizontal $U(1)$ symmetries in the  observable sector correspond to
$U(1)_j$ $(j=1,\cdots,3)$ and to the world--sheet current
${\bar\omega}^2{\bar\omega}^3$.
The $U(1)$ symmetries in the
hidden sector, $U(1)_7$ and $U(1)_8$,
correspond to the world--sheet currents
${\bar\phi}^1{\bar\phi}^{1^*}+{\bar\phi}^8{\bar\phi}^{8^*}$ and
$-2{\bar\phi}^j{\bar\phi}^{j^*}+{\bar\phi}^1{\bar\phi}^{1^*}
-4{\bar\phi}^2{\bar\phi}^{2^*}-{\bar\phi}^8{\bar\phi}^{8^*}$ respectively,
where summation on $j=5,\cdots,7$ is implied.

The massless spectrum in the observable sector
contains three chiral generations
from the sectors $b_1$, $b_2$ and $b_3$,
$G_{1_{{1\over2},0,0,0}}+G_{2_{0,{1\over2},0,0}}+
\{({e^c}+{u^c})_{0,0,{1\over2},-{1\over2}}+
({d^c}+{N^c})_{0,0,{1\over2},{1\over2}}+
(L)_{0,0,{1\over2},-{1\over2}}+(Q)_{0,0,{1\over2},{1\over2}}\}_{_3}.$
The Neveu--Schwarz
sector contains in  addition to the spin two and spin one
states, one pair of Higgs doublets $h_{3_{0,0,1}}$,
${\bar h}_{3_{0,0,-1}}$,
two pairs of Higgs triplets $D_{1_{-1,0,0}}$,
${\bar D}_{1_{1,0,0}}$, $D_{2_{0,-1,0}}$,
${\bar D}_{2_{0,1,0}}$, one pair of $SO(10)$ singlets with
charges under the horizontal $U(1)$ symmetries,
$\Phi_{{12}_{1,-1,0,0}}$,
${\bar\Phi}_{{12}_{-1,1,0,0}}$ and five singlets which are
neutral under all the $U(1)$ symmetries
$\xi_{1,\cdots,5}:{\hskip .2cm}
{\chi^{12}_{1\over2}{\bar y}^1_{1\over2}
{\bar\omega}^1_{1\over2}{\vert 0\rangle}_0},$
 ${\chi^{34}_{{1\over2}}{\bar y}_{1\over2}^4{\bar\omega}_{1\over2}^4
{\vert 0\rangle}_0},$
 $\chi^{56}_{1\over2}{\bar y}_{1\over2}^2{\bar y}_{1\over2}^3
{\vert 0\rangle}_0$,
$\chi^{56}_{1\over2}{\bar y}_{1\over2}^5{\bar\omega}_{1\over2}^5
{\vert 0\rangle}_0$ and
$\chi^{56}_{1\over2}{\bar y}_{1\over2}^6{\bar\omega}_{1\over2}^6
{\vert 0\rangle}_0.$

In addition, in the observable sector,
the sector $\zeta=\alpha+\beta$ gives
$$\eqalignno{h_{45}&\equiv{[(1,0);(2,-1)]}_
{{1\over2},{1\over2},0,0} {\hskip .5cm}
D_{45}\equiv{[(3,-1);(1,0)]}_
{{1\over2},{1\over2},0,0}&(15a,b)\cr
\Phi_{45}&\equiv{[(1,0);(1,0)]}_
{{1\over2},{1\over2},-1,0}  {\hskip .8cm}
\Phi^{\pm}_3\equiv{[(1,0);(1,0)]}_
{-{1\over2},{1\over2},0,\pm1}&(15e,f)\cr
\phi_1,\phi^\prime_1&\equiv{[(1,0);(1,0)]}_
{-{1\over2},{1\over2},0,0} {\hskip .5cm}
\phi_2,\phi^\prime_2\equiv{[(1,0);(1,0)]}_
{-{1\over2},{1\over2},0,0}&(15e,f)}$$
(and their conjugates ${\bar h}_{45}$, etc.).
The states are obtained by acting on the vacuum
with the fermionic oscillators
${\bar\psi}^{4,5},{\bar\psi}^{1,...,3},{\bar\eta}^3,{\bar\omega}^2\pm
i{\bar\omega}^3,{\bar y}^5,{\bar y}^6,
{\bar\omega}^5,{\bar\omega}^6$,
respectively  (and their complex conjugates for ${\bar h}_{45}$, etc.).

The sectors $b_i+2\gamma+(I){\hskip .2cm} (i=1,..,3)$ give vector
representations which are
$SU(3)_C\times SU(2)_L\times {U(1)_L}\times {U(1)_C}$
singlets (see Table 6). The vectors with some combination
of $(b_1,b_2,b_3,\alpha,\beta)$
plus $\gamma+(I)$ (see Table 7) give  representations which transform
under $SU(3)_C\times SU(2)_L\times {U(1)_L}\times {U(1)_C}$, most of
them singlets, but carry either
$U(1)_Y$ or $U(1)_{Z^\prime}$ charges. Some of these states carry
fractional charges $\pm{1\over2}$ or $\pm{1\over3}$.
There are no representations that transform nontrivially both under the
observable
and hidden sectors. The only mixing which occurs is of states that transform
nontrivially under the observable or hidden sectors and carry U(1) charges
under the hidden or observable sectors respectively.

The model contains eight $U(1)$ symmetries, six in the observable
sector and two in the hidden sector. Out of those four are anomaly free
and four are anomalous:

\parindent=15pt

Tr${U_1}=18$, Tr${U_2}=30$, Tr${U_3}=24$,
Tr${U_4}=12$.
{\hskip 5cm}(16)

\parindent=15pt

The two trace $U(1)$s, $U(1)_L$ and $U(1)_C$, are anomaly free.
Consequently, the weak hypercharge and the orthogonal combination,
$U(1)_{Z^{\prime}}$, are
anomaly free. Likewise, the two $U(1)$s in the hidden sector are anomaly free.
Of the four anomalous $U(1)$s, only three can be rotated by
an orthogonal transformation and one combination remains anomalous and is
uniquely given by: ${U_A}=k{\sum_j} [{Tr {U(1)_j}}]U(1)_j$,
where $j$ runs over all the
anomalous $U(1)$s.
For convenience, I take $k={1\over6}$, and therefore
the anomalous combination
is given by:
$$U_A=3U_1+5U_2+4U_3+2U_4,{\hskip 1cm}TrQ_A=318.\eqno(17a)$$
The three orthogonal combinations are not unique. Different
choices are related by orthogonal transformations. One choice is given by:
$$\eqalignno{{U^\prime}_1&=U_1+U_2-2U_3&(17b)\cr
{U^\prime}_2&=U_1-U_2+U_4&(17c)\cr
{U^\prime}_3&=3U_1-U_2+U_3-4U_4.&(17d)\cr}$$
Together with the other four anomaly free $U(1)$s,
they are free from  gauge and gravitational
anomalies. The cancellation of all mixed  anomalies among the five $U(1)$s
is a non trivial consistency check of the
massless spectrum of the model.

The trilevel superpotential is given by
$$\eqalignno{W&=\{(
{u_{L_1}^c}{e_{L_1}^c}{D}_1+{d_{L_1}^c}{N_{L_1}^c}{D}_1+
{u_{L_2}^c}{e_{L_2}^c}{D}_2+{d_{L_1}^c}{N_{L_2}^c}{D}_1+
{u_{L_3}^c}Q_3{\bar h}_3+{N_{L_3}^c}L_3{\bar h}_3)\cr
&\qquad
+{{D_1}{\bar D}_2{\bar\Phi}_{12}}
+{\bar D}_1{D}_2{\Phi}_{12}
+{\bar\Phi}_{12}{\bar\Phi}_3^+{\bar\Phi}_3^-
+{\Phi_{12}}\Phi_3^-\Phi_3^+
+h_3{\bar h}_{45}\Phi_{45}+
{\bar h}_3h_{45}{\bar\Phi}_{45}\cr
&\qquad
+{1\over2}\xi_3(\Phi_{45}{\bar\Phi}_{45}+h_{45}{\bar h}_{45}
+D_{45}{\bar D}_{45}+\phi_1{\bar\phi}_1+\phi_1^\prime{\bar\phi}_1^\prime+
\phi_2{\bar\phi}_2+\phi_2^\prime{\bar\phi}_2^\prime
+\Phi_3^+{\bar\Phi}_3^+\cr
&\qquad
+\Phi_3^-{\bar\Phi}_3^-+H_1H_2)+
\phi_1(M_3M_{11}+M_2M_9)+{\bar\phi}_1M_6M_{13}+
{\bar\phi}_2^\prime(M_4M_{10}\cr
&\qquad
+M_{5}M_{12})+\phi_2^\prime(
M_7M_{14}+M_1M_8)+\phi^\prime_1M_{17}M_{24}
+{\bar\phi}^\prime_1(M_{16}M_{21}+M_{20}M_{23})\cr
&\qquad
+{\bar\phi}_2(M_{15}M_{22}+M_{19}M_{26})
+\phi_2M_{18}M_{25}+{\bar\Phi}_{12}H_{13}H_{14},\quad&(18)\cr}$$
where a common normalization constant ${\sqrt 2}g$ is assumed.

The solutions to Eqs. (14) can be divided to two kinds of solutions.
Solutions of the first kind
keep   both $U(1)_C$ and $U(1)_L$ unbroken. Solutions of the
second kind keep only the
electroweak hypercharge unbroken. Solutions of the first kind are
preferred because they are believed to be stable to all orders.
 For the first kind of  solutions
the fields  $\chi_k$ in Eqs. (14), must be  neutral under
both $U(1)_C$ and $U(1)_L$.
Only the Neveu--Schwarz sector, the sector
$\zeta$, and the sectors $b_j+2\gamma$, produce fields which are
neutral under both $U(1)_C$ and $U(1)_L$.
By examining the massless states from the Neveu--Schwarz and the  $\zeta$
sector, it is observed that the number of fields, with independent
charges along the four D constraints is always less than four.
The Neveu Schwarz sector produces only one field, $\Phi_{12}$. The sector
$\zeta$ gives $\Phi_{45}$ and $\Phi_3^\pm$ while $\phi_{1,2}$ and
$\phi^\prime_{1,2}$ have the same charges, up to a multiplicative
constant, as $\Phi_{12}$. However only three of the four fields
have independent charges. The complex conjugate fields can be
used to relax the positive definite restriction, however do not add
more degrees of freedom.
Thus
the number of constraints is larger than the number of fields which
can be used to solve theim.
Adding the states from the sectors $b_j+2\gamma$ does not resolve
the problem, since they always carry positive charge along the
anomalous $U(1)_A$\footnote*{From the modular invariance rules,
it can be shown that
the states in the sectors $b_j$ and $b_j+2\gamma$ carry identical charges
along $U(1)_j$. The states from these sectors determine the sign of the
anomaly and therefore have positive $U(1)_A$ charges}.
 Changing the model to include more states from
the Neveu--Schwarz sector is possible at the cost of increasing the
number of $U(1)$ symmetries with non vanishing trace. Thus, it is
found that the number of constraints is always larger than the number
of flat directions. It is concluded that
solutions of the first kind do not exist in type {\bf {I}} models.
This result was verified by  writing a simple computer program
which searches for positive definite solutions. No solutions were
found in all type {\bf I} models.
It is therefore concluded that, solutions which keep both
$U(1)_C$ and $U(1)_L$ unbroken
by the Dine--Seiberg--Witten mechanism, do not exist
in type {\bf I} models.

Turning to the second kind of solutions.
These solutions  keep only the weak
hypercharge unbroken in the application of the Dine--Seiberg--Witten
mechanism. The set of fields which can receive a non vanishing
VEV is extended to include the states with vanishing weak hypercharge,
but with non vanishing $U(1)_{Z^\prime}$ charge.
These states include the three right handed neutrinos
from the sectors $b_1$, $b_2$ and $b_3$, and the neutral
states  from the sectors  $\pm\gamma+(I)$ plus some combination
of $(b_1,b_2,b_3,\alpha,\beta)$ (see Table 7).
The number of D flatness constraints in this case is extended to ten
equations. To obtain a supersymmetric vacuum we take
 $W={{\partial W}\over{\partial{\eta_i}}}=0$, where $W$ is the
trilevel superpotential.
An elaborate computerized search for F and D flat
solutions yielded a null result. However at this stage it is not possible
to present a definite conclusion whether solutions of the second kind
exist or do not exist in type {\bf I} models. Observation of an additional
neutral gauge boson, $Z^\prime$, will exclude this kind of solutions
and will therefore exclude type {\bf I} models.

There is a unique
class of type {\bf II}  models [\EU,\TOP]
which admit  solutions to the F and D flatness constraints.
These  models have the following characteristics:

\parindent=-15pt

1.  The boundary condition vectors allow a trilevel Yukawa coupling
only for  $+{2\over3}$ charged quarks.
The mass terms for the lighter quarks and leptons are obtained
from nonrenormalizable terms.

2. The complexification of the left--moving fermions $y^3y^6$, $y^1\omega^5$
and $\omega^2\omega^4$ allows the construction of a vector $\zeta$.
The states from this sector are used in the application of the DSW
mechanism.

3. These models are constructed at a highly symmetric point in the
``compactified space". This symmetry exhibits itself in the
non vanishing $U(1)$ traces [\EU,\TOP].

\parindent=15pt

\bigskip
\centerline{\bf 5.  The  Hidden Sector}

The hidden sector in the free fermionic standard--like  models is determined
by the boundary condition of the internal right--moving fermions,
${\bar\phi}^{1,\cdots,8}$.  A detailed classification is beyond the
scope of this paper. However, the following comments are important to note.

The hidden gauge group arises from the states ${\bar\phi}^{1,\cdots,8}$.
In the NAHE set the contribution to the hidden $E_8$
gauge group  comes from the Neveu--Schwarz sector
and from the sector $I=1+b_1+b_2+b_3$. In the standard--like models
the hidden gauge group is broken
by the vectors which extend the NAHE set.

It is important to note that in the standard--like models the hidden
$E_8$ gauge
group must be broken. This follows from the
fact that the vectors which break the $SO(10)$ symmetry always
carry an odd number of periodic fermions from the set
$\{{\bar\psi}^{1,\cdots5},{\bar\eta}^1,{\bar\eta}^2,{\bar\eta}^3\}$.
The  reason is the structure of the NAHE set, which divides the
internal fermionic states into three symmetric groups and the
requirement of at least one Higgs doublet from the Neveu Schwarz sector.
To obey the modular invariance rule, $\alpha\cdot\gamma=0mod1$,
an odd number of fermions from the set $\{{\bar\phi}^{1,\cdots,8}\}$
must be periodic in the vector $\alpha$, and receive  boundary
condition of $1\over2$
in the vector $\gamma$. Therefore, the hidden gauge symmetry
is  broken in two stages. Typically it is broken to $SU(5)\times
SU(3)\times U(1)^2$. However, other possibilities do exist.

In the free fermionic models the hidden sector contains a rich
matter spectrum. The hidden
matter spectrum comes in vector representations
Therefore we may expect matter condensates to form at the scale
in which the
hidden symmetries become strong. In the standard--like models
small gauge groups, like $SU(3)$, usually appear. This offers
the possibility of a rich hidden
matter spectrum to appear in future colliders.
The appearance of small hidden gauge groups may be desirable for
generating the breaking of $U(1)_{Z^\prime}$ as well as for generating
supersymmetry breaking at a low scale.

\bigskip
\centerline{\bf 6.  Discussion}

The construction of free fermionic standard--like models led to a unique
class of models. This class of models has unique phenomenological
characteristics. They suggest an explanation for the
top--bottom mass hierarchy. At the trilinear level of the superpotential
only the top quark obtains a non vanishing mass term, while the lighter
quarks and leptons get their
mass terms from non renormalizable terms. In two recent constructions
[\EU,\TOP], mass
terms for the bottom quark and for the tau lepton were found at the
quartic and quintic level. These models predict a top quark at
$m_t\sim140-180GeV$ [\YUKAWA].
The unsuccessful search for solutions to the F
and D flatness constraints in type {\bf I} models, suggests  a possible
connection between the requirement of a supersymmetric vacuum at
the Planck scale and the top quark mass
hierarchy. Observation of an additional neutral gauge boson
$Z^\prime$ will be further evidence to support this connection.

The standard--like models extend the symmetry of the Standard--Model
by one additional, generation independent,  $U(1)$ symmetry. This
$U(1)$ symmetry is a combination of, $B-L$, baryon minus lepton number,
and of $T_{3_R}$. The $U(1)_{Z^\prime}$ may be broken by the application
of the DSW mechanism. However, if $U(1)_{Z^\prime}$ remains unbroken
down to $M_{Z^\prime}\le{10}^7GeV$, it results in a gauged mechanism to
explain the suppression of proton decay from
dimension four operators. In this case it may be
broken by the running of the
renormalization group equations, {\`a} la no--scale supergravity.
Another possibility is that it is broken by a condensate of the hidden
gauge group with non vanishing $U(1)_{Z^\prime}$ charge. The presence of small
gauge groups like $SU(3)$ makes this solution very appealing.
Another possible consequence of
matter condensates in the hidden gauge group is the breaking of global
supersymmetry [\AEHN].

The underlying $SO(10)$ symmetry of the NAHE set indicates
that for every Dirac mass term for a $+{2\over3}$ charged quark,
we obtain a Dirac mass term for a neutral lepton, with $m_u=m_\nu$.
Therefore, we must be able to
construct a see--saw mechanism [\NMASSES] to suppress the neutrino mass.
The entries in the see--saw mass matrix arise from nonrenormalizable
terms. For example, in the models of Refs. [\EU,\TOP] a potential term
in the see--saw mass matrix appears at the quartic level
$N_L^cH_{17}H_{13}V_9$, where
$V_9$ and $H_{13}$ transform as triplets under the
hidden $SU(3)$ group.

In this paper I discussed the construction of superstring standard--like
models in the free fermionic formulation. To date the free fermionic
formulation yielded the most realistic superstring models.
This realism may be not accidental but may arise from the
fact that the  free fermionic formulation is formulated at a highly symmetric
point in the moduli space. The question, how does nature choose to have
only three generations, finds a simple explanation in free fermionic
models [\naturalness].
The free fermionic standard--like models have remarkable properties.
They have exactly three generations and no mirror
generations. They explain the suppression of proton decay via  dimension
four operators either by a gauged mechanism or by simply not giving a VEV
to the neutral singlet in the $10$ of $SU(5)$. They explain the
suppression of proton decay via dimension five operators by the GSO
projection of the dangerous Higgs triplets. The projection of the
Higgs triplets is correlated with the appearance of horizontal
$U(1)_{\ell,r}$ symmetries. The standard--like models suggest an
explanation for one of the most important mysteries  of nature,
the heaviness of the top quark relative to the lighter quarks and leptons.
At trilevel only the top quark obtains a non vanishing mass term.
Therefore only the top quark mass is characterized by the
electroweak scale. The mass terms for the lighter quarks and leptons are
obtained from nonrenormalizable terms and therefore are
naturally suppressed.
Preliminary studies of nonrenormalizable terms in these models
indicate that construction of realistic mass matrices is possible
and will be reported elsewhere.
To conclude, we have made the first steps on the way
toward a superstring standard model.

\bigskip
\centerline{\bf Acknowledgments}
I thank Aviva Faraggi and Doron Gepner for
useful discussions. This work is supported in part by a Feinberg
School Fellowship.

\refout

\vfill
\eject

\input tables.tex

\hoffset=1.5truein
\nopagenumbers
\magnification=1000
\font\normalroman=cmr10
\font\style=cmr7
\style

\fontdimen12\fivesy=0pt

\textfont0=\sevenrm
\scriptfont0=\fiverm
\textfont1=\seveni
\scriptfont1=\fivei
\textfont2=\sevensy
\scriptfont2=\fivesy

{\hfill
{\begintable
\  \ \|\ $\psi^\mu$ \ \|\ $\{\chi^{12};\chi^{34};\chi^{56}\}$ \ \|\
$y^3{\bar y}^3$,  $y^4{\bar y}^4$, $y^5{\bar y}^5$,  ${y}^6{\bar y}^6$
\ \|\ $y^1{\bar y}^1$,  $y^2{\bar y}^2$,  $\omega^5{\bar\omega}^5$,
${\omega}^6{\bar\omega}^6$
\ \|\ $\omega^1{\bar\omega}^1$,  $\omega^2{\bar\omega}^2$,
$\omega^3{\bar\omega}^3$,  ${\omega}^4{\bar\omega}^4$ \ \|\
${\bar\psi}^1$, ${\bar\psi}^2$, ${\bar\psi}^3$,
${\bar\psi}^4$, ${\bar\psi}^5$, ${\bar\eta}^1$,
${\bar\eta}^2$, ${\bar\eta}^3$ \ \|\
${\bar\phi}^1$, ${\bar\phi}^2$, ${\bar\phi}^3$, ${\bar\phi}^4$,
${\bar\phi}^5$, ${\bar\phi}^6$, ${\bar\phi}^7$, ${\bar\phi}^8$ \crthick
$b_4$
\|\ 1 \| $\{1,~0,~0\}$ \|
1, ~~~0, ~~~~0, ~~~~1 \|
0, ~~~0, ~~~~1, ~~~~0 \|
0, ~~~0, ~~~~0, ~~~~1 \|
1, ~~1, ~~1, ~~0, ~~0, ~~1, ~~0, ~~0 \|
1, ~~1, ~~0, ~~0, ~~0, ~~0, ~~0, ~~0 \nr
$\alpha$
\|\ 1 \| $\{0,~1,~0\}$ \|
0, ~~~0, ~~~~0, ~~~~1 \|
0, ~~~1, ~~~~1, ~~~~0 \|
1, ~~~0, ~~~~0, ~~~~0 \|
1, ~~1, ~~1, ~~0, ~~0, ~~0, ~~1, ~~0 \|
1, ~~1, ~~0, ~~0, ~~0, ~~0, ~~0, ~~0 \nr
$\gamma$
\|\ 1 \| $\{0,~0,~1\}$ \|
1, ~~~1, ~~~~0, ~~~~0 \|
1, ~~~0, ~~~~0, ~~~~0 \|
0, ~~~1, ~~~~0 ,~~~~0 \|
$1\over2$, ~~$1\over2$, ~~$1\over2$, ~~$1\over2$,
{}~~$1\over2$, ~~$1\over2$, ~~$1\over2$, ~~$1\over2$, \|
$1\over2$, ~~$1\over2$,
{}~~$1\over2$, ~~$1\over2$, ~~1, ~~0, ~~0, ~~0 \endtable}
\hfill}
\bigskip
\parindent=0pt
\hangindent=39pt\hangafter=1
\normalroman

{\it Table 1.} A three generations $SU(3)\times SU(2)\times U(1)^2$ model
without Higgs doublets from the Neveu--Schwarz sector.
\vskip 2.5cm

\input tables.tex

\hoffset=1.5truein
\nopagenumbers
\magnification=1000
\font\normalroman=cmr10
\font\style=cmr7
\style

\fontdimen12\fivesy=0pt

\textfont0=\sevenrm
\scriptfont0=\fiverm
\textfont1=\seveni
\scriptfont1=\fivei
\textfont2=\sevensy
\scriptfont2=\fivesy

{\hfill
{\begintable
\  \ \|\ $\psi^\mu$ \ \|\ $\{\chi^{12};\chi^{34};\chi^{56}\}$ \ \|\
$y^3y^6$,  $y^4{\bar y}^4$, $y^5{\bar y}^5$,  ${\bar y}^3{\bar y}^6$
\ \|\ $y^1\omega^6$,  $y^2{\bar y}^2$,  $\omega^5{\bar\omega}^5$,
${\bar y}^1{\bar\omega}^6$
\ \|\ $\omega^1{\omega}^3$,  $\omega^2{\bar\omega}^2$,
$\omega^4{\bar\omega}^4$,  ${\bar\omega}^1{\bar\omega}^3$ \ \|\
${\bar\psi}^1$, ${\bar\psi}^2$, ${\bar\psi}^3$,
${\bar\psi}^4$, ${\bar\psi}^5$, ${\bar\eta}^1$,
${\bar\eta}^2$, ${\bar\eta}^3$ \ \|\
${\bar\phi}^1$, ${\bar\phi}^2$, ${\bar\phi}^3$, ${\bar\phi}^4$,
${\bar\phi}^5$, ${\bar\phi}^6$, ${\bar\phi}^7$, ${\bar\phi}^8$ \crthick
${\alpha}$
\|\ 1 \| $\{1,~0,~0\}$ \|
1, ~~~0, ~~~~0, ~~~~1 \|
0, ~~~0, ~~~~1, ~~~~0 \|
0, ~~~0, ~~~~1, ~~~~0 \|
1, ~~1, ~~1, ~~1, ~~1, ~~1, ~~0, ~~0 \|
0, ~~0, ~~0, ~~0, ~~0, ~~0, ~~0, ~~0 \nr
$\beta$
\|\ 1 \|
$\{0,~0,~1\}$ \|
0, ~~~0, ~~~~0, ~~~~1 \|
0, ~~~1, ~~~~0, ~~~~1 \|
1, ~~~0, ~~~~1, ~~~~0 \|
1, ~~1, ~~1, ~~0, ~~0, ~~1, ~~0, ~~1 \|
1, ~~1, ~~1, ~~1, ~~0, ~~0, ~~0, ~~0 \nr
$\gamma$
\|\ 1 \| $\{0,~1,~0\}$ \|
0, ~~~0, ~~~~1, ~~~~1 \|\
1, ~~~0, ~~~~0, ~~~~1 \|
0, ~~~1, ~~~~0, ~~~~0 \|
{}~~$1\over2$, ~~$1\over2$, ~~$1\over2$, ~~$1\over2$,
{}~~$1\over2$, ~~$1\over2$, ~~$1\over2$, ~~$1\over2$, \|
 $1\over2$, ~~0, ~~1, ~~1,
{}~~$1\over2$,
{}~~$1\over2$, ~~$1\over2$, ~~0 \endtable}
\hfill}
\smallskip
\parindent=0pt
\hangindent=39pt\hangafter=1
\normalroman

{{\it Table 2.} A three generations $SU(3)\times SU(2)\times U(1)^2$ model.
The choice of generalized GSO coeficients is:
$c\left(\matrix{{\alpha}\cr
                                    b_j,\beta\cr}\right)=
-c\left(\matrix{{\alpha}\cr
                                    1\cr}\right)=
c\left(\matrix{\beta\cr
                                    1\cr}\right)=
c\left(\matrix{\beta\cr
                                    b_j\cr}\right)=
-c\left(\matrix{\beta\cr
                                    \gamma\cr}\right)=
c\left(\matrix{\gamma\cr
                                    b_2\cr}\right)=
-c\left(\matrix{\gamma\cr
                                    b_1,b_3,{\alpha},\gamma\cr}\right)=
-1$ (j=1,2,3), with the others specified by modular invariance and space--time
supersymmetry.
Trilevel Yukawa couplings are obtained for $+{2\over3}$ charged quarks
 as well as $-{1\over3}$ charged quarks and for charged leptons.
\vskip 1.5cm

\vfill
\eject

\input tables.tex

\hoffset=1.5truein
\nopagenumbers
\magnification=1000
\font\normalroman=cmr10
\font\style=cmr7
\style

\fontdimen12\fivesy=0pt

\textfont0=\sevenrm
\scriptfont0=\fiverm
\textfont1=\seveni
\scriptfont1=\fivei
\textfont2=\sevensy
\scriptfont2=\fivesy

{\hfill
{\begintable
\  \ \|\ $\psi^\mu$ \ \|\ $\{\chi^{12};\chi^{34};\chi^{56}\}$ \ \|\
$y^3y^6$,  $y^4{\bar y}^4$, $y^5{\bar y}^5$,  ${\bar y}^3{\bar y}^6$
\ \|\ $y^1\omega^5$,  $y^2{\bar y}^2$,  $\omega^6{\bar\omega}^6$,
${\bar y}^1{\bar\omega}^5$
\ \|\ $\omega^2{\omega}^4$,  $\omega^1{\bar\omega}^1$,
$\omega^3{\bar\omega}^3$,  ${\bar\omega}^2{\bar\omega}^4$ \ \|\
${\bar\psi}^1$, ${\bar\psi}^2$, ${\bar\psi}^3$,
${\bar\psi}^4$, ${\bar\psi}^5$, ${\bar\eta}^1$,
${\bar\eta}^2$, ${\bar\eta}^3$ \ \|\
${\bar\phi}^1$, ${\bar\phi}^2$, ${\bar\phi}^3$, ${\bar\phi}^4$,
${\bar\phi}^5$, ${\bar\phi}^6$, ${\bar\phi}^7$, ${\bar\phi}^8$ \crthick
$\alpha$
\|\ 0 \|
$\{0,~0,~0\}$ \|
1, ~~~0, ~~~~0, ~~~~0 \|
0, ~~~0, ~~~~1, ~~~~1 \|
0, ~~~0, ~~~~1, ~~~~1 \|
1, ~~1, ~~1, ~~0, ~~0, ~~0 ,~~0, ~~0 \|
1, ~~1, ~~1, ~~1, ~~0, ~~0, ~~0, ~~0 \nr
$\beta$
\|\ 0 \| $\{0,~0,~0\}$ \|
0, ~~~0, ~~~~1, ~~~~1 \|
1, ~~~0, ~~~~0, ~~~~0 \|
0, ~~~1, ~~~~0, ~~~~1 \|
1, ~~1, ~~1, ~~0, ~~0, ~~0, ~~0, ~~0 \|
1, ~~1, ~~1, ~~1, ~~0, ~~0, ~~0, ~~0 \nr
$\gamma$
\|\ 0 \|
$\{0,~0,~0\}$ \|
0, ~~~1, ~~~~0, ~~~~1 \|\
0, ~~~1, ~~~~0, ~~~~1 \|
1, ~~~0, ~~~~0, ~~~~0 \|
{}~~$1\over2$, ~~$1\over2$, ~~$1\over2$, ~~$1\over2$,
{}~~$1\over2$, ~~$1\over2$, ~~$1\over2$, ~~$1\over2$ \| $1\over2$, ~~0, ~~1,
{}~~1,
{}~~$1\over2$,
{}~~$1\over2$, ~~$1\over2$, ~~0 \endtable}
\hfill}
\smallskip
\parindent=0pt
\hangindent=39pt\hangafter=1
\normalroman

{{\it Table 3.} A three generations $SU(3)\times SU(2)\times U(1)^2$ model.
The choice of generalized GSO coeficients is:
$c\left(\matrix{b_j\cr
                                    \alpha,\beta,\gamma\cr}\right)=
-c\left(\matrix{\alpha\cr
                                    1\cr}\right)=
c\left(\matrix{\alpha\cr
                                    \beta\cr}\right)=
-c\left(\matrix{\beta\cr
                                    1\cr}\right)=
c\left(\matrix{\gamma\cr
                                    1,\alpha\cr}\right)=
-c\left(\matrix{\gamma\cr
                                    \beta\cr}\right)=
-1$ (j=1,2,3), with the others specified by modular invariance and space--time
supersymmetry.
Trilevel Yukawa couplings are obtained only for $+{2\over3}$ charged quarks. }
\vskip 2.5cm

\input tables.tex

\hoffset=1.5truein
\nopagenumbers
\magnification=1000
\font\normalroman=cmr10
\font\style=cmr7
\style

\fontdimen12\fivesy=0pt

\textfont0=\sevenrm
\scriptfont0=\fiverm
\textfont1=\seveni
\scriptfont1=\fivei
\textfont2=\sevensy
\scriptfont2=\fivesy

{\hfill
{\begintable
\  \ \|\ $\psi^\mu$ \ \|\ $\{\chi^1;\chi^2;\chi^3\}$ \ \|\
$y^3y^6$,  $y^4{\bar y}^4$, $y^5{\bar y}^5$,  ${\bar y}^3{\bar y}^6$
\ \|\ $y^1\omega^5$,  $y^2{\bar y}^2$,  $\omega^6{\bar\omega}^6$,
${\bar y}^1{\bar\omega}^5$
\ \|\ $\omega^2{\omega}^4$,  $\omega^1{\bar\omega}^1$,
$\omega^3{\bar\omega}^3$,  ${\bar\omega}^2{\bar\omega}^4$ \ \|\
${\bar\psi}^1$, ${\bar\psi}^2$, ${\bar\psi}^3$,
${\bar\psi}^4$, ${\bar\psi}^5$, ${\bar\eta}^1$,
${\bar\eta}^2$, ${\bar\eta}^3$ \ \|\
${\bar\phi}^1$, ${\bar\phi}^2$, ${\bar\phi}^3$, ${\bar\phi}^4$,
${\bar\phi}^5$, ${\bar\phi}^6$, ${\bar\phi}^7$, ${\bar\phi}^8$ \crthick
$\alpha$
\|\ 0 \|
$\{0,~0,~0\}$ \|
1, ~~~1, ~~~~1, ~~~~0 \|
1, ~~~1, ~~~~1, ~~~~0 \|
1, ~~~1, ~~~~1, ~~~~0 \|
1, ~~1, ~~1, ~~0, ~~0, ~~0, ~~0, ~~0 \|
1, ~~1, ~~1, ~~1, ~~0, ~~0, ~~0, ~~0 \nr
$\beta$
\|\ 0 \| $\{0,~0,~0\}$ \|
0, ~~~1, ~~~~0, ~~~~1 \|
0, ~~~1, ~~~~0, ~~~~1 \|
1, ~~~0, ~~~~0, ~~~~0 \|
1, ~~1, ~~1, ~~0, ~~0, ~~0, ~~0, ~~0 \|
1, ~~1, ~~1, ~~1, ~~0, ~~0, ~~0, ~~0 \nr
$\gamma$
\|\ 0 \|
$\{0,~0,~0\}$ \|
0, ~~~0, ~~~~1, ~~~~1 \|\
1, ~~~0, ~~~~0, ~~~~0 \|
0, ~~~1, ~~~~0, ~~~~1 \|
 ~~$1\over2$, ~~$1\over2$, ~~$1\over2$, ~~$1\over2$,
{}~~$1\over2$, ~~$1\over2$, ~~$1\over2$, ~~$1\over2$ \| $1\over2$, ~~0, ~~1,
{}~~1,
{}~~$1\over2$,
{}~~$1\over2$, ~~$1\over2$, ~~0 \endtable}
\hfill}
\smallskip
\parindent=0pt
\hangindent=39pt\hangafter=1
\normalroman

{{\it Table 4.} A three generations $SU(3)\times SU(2)\times U(1)^2$ model.
The choice of generalized GSO coeficients is:
$c\left(\matrix{b_j\cr
                                    \alpha,\beta,\gamma\cr}\right)=
-c\left(\matrix{\alpha\cr
                                    1\cr}\right)=
-c\left(\matrix{\alpha\cr
                                    \beta\cr}\right)=
-c\left(\matrix{\beta\cr
                                    1\cr}\right)=
c\left(\matrix{\gamma\cr
                                    1\cr}\right)=
-c\left(\matrix{\gamma\cr
                                   \alpha,\beta\cr}\right)=
-1$ (j=1,2,3), with the others specified by modular invariance and space--time
supersymmetry.
Trilevel Yukawa couplings are obtained only for $+{2\over3}$ charged quarks. }
\vskip 2cm

\vfill
\eject

\input tables.tex

\hoffset=1.5truein
\nopagenumbers
\magnification=1000
\font\normalroman=cmr10
\font\style=cmr7
\style

\fontdimen12\fivesy=0pt

\textfont0=\sevenrm
\scriptfont0=\fiverm
\textfont1=\seveni
\scriptfont1=\fivei
\textfont2=\sevensy
\scriptfont2=\fivesy

{\hfill
{\begintable
\  \ \|\ $\psi^\mu$ \ \|\ $\{\chi^{12};\chi^{34};\chi^{56}\}$ \ \|\
$y^3{\bar y}^3$,  $y^4{\bar y}^4$, $y^5{\bar y}^5$,  ${y}^6{\bar y}^6$
\ \|\ $y^1{\bar y}^1$,  $y^2{\bar y}^2$,  $\omega^5{\bar\omega}^5$,
${\omega}^6{\bar\omega}^6$
\ \|\ $\omega^2{\omega}^3$,  $\omega^1{\bar\omega}^1$,
$\omega^4{\bar\omega}^4$,  ${\bar\omega}^2{\bar\omega}^3$ \ \|\
${\bar\psi}^1$, ${\bar\psi}^2$, ${\bar\psi}^3$,
${\bar\psi}^4$, ${\bar\psi}^5$, ${\bar\eta}^1$,
${\bar\eta}^2$, ${\bar\eta}^3$ \ \|\
${\bar\phi}^1$, ${\bar\phi}^2$, ${\bar\phi}^3$, ${\bar\phi}^4$,
${\bar\phi}^5$, ${\bar\phi}^6$, ${\bar\phi}^7$, ${\bar\phi}^8$ \crthick
$\alpha$
\|\ 1 \|
$\{1,~0,~0\}$ \|
1, ~~~0, ~~~~0, ~~~~1 \|
0, ~~~0, ~~~~1, ~~~~0 \|
0, ~~~0, ~~~~1, ~~~~1 \|
1, ~~1, ~~1, ~~0, ~~0, ~~1, ~~0, ~~1\|
1, ~~1, ~~1, ~~1, ~~0, ~~0, ~~0, ~~0 \nr
$\beta$
 \|\ 1 \|
$\{0,~1,~0\}$ \|
0, ~~~0, ~~~~0, ~~~~1 \|
0, ~~~1, ~~~~1, ~~~~0 \|
0, ~~~1, ~~~~0, ~~~~1 \|
1, ~~1, ~~1, ~~0, ~~0, ~~0, ~~1, ~~1 \|
1, ~~1, ~~1, ~~1, ~~0, ~~0, ~~0, ~~0 \nr
$\gamma$
 \|\ 1 \|
$\{0,~0,~1\}$ \|
1, ~~~1, ~~~~0, ~~~~0 \|
1, ~~~1, ~~~~0, ~~~~0 \|
0, ~~~0, ~~~~0, ~~~~1 \|
$1\over2$, ~~$1\over2$, ~~$1\over2$, ~~$1\over2$,
{}~~$1\over2$, ~~$1\over2$, ~~$1\over2$, ~~$1\over2$, \| $1\over2$, ~~0
{}~~1, ~~1, ~~$1\over2$, ~~$1\over2$, ~~$1\over2$, ~~0 \endtable}
\hfill}
\bigskip
\parindent=0pt
\hangindent=39pt\hangafter=1
\normalroman

{\it Table 5.} A three generations $SU(3)\times SU(2)\times U(1)^2$ model
with four horizontal $U(1)$ symmetries.
The choice of generalized GSO coeficients is:
$c\left(\matrix{b_1,b_3,\alpha,\beta,\gamma\cr
                                    \alpha\cr}\right)=
-c\left(\matrix{b_2\cr
                                    \alpha\cr}\right)=
c\left(\matrix{1,b_j,\gamma\cr
                                    \beta\cr}\right)=
-c\left(\matrix{\gamma\cr
                                    1,b_1,b_2\cr}\right)=
c\left(\matrix{\gamma\cr
                                   b_3\cr}\right)=
-1$ (j=1,2,3), with the others specified by modular invariance and space--time
supersymmetry.
Trilevel Yukawa couplings are obtained for $+{2\over3}$ charged quarks
as well as for $-{1\over3}$ charged quarks and for charged leptons. }

\vfill
\eject

\input tables.tex
\nopagenumbers
\magnification=1000
\baselineskip=18pt
\hbox
{\hfill
{\begintable
\ F \ \|\ SEC \ \|\ $SU(3)_C$ $\times$ $SU(2)_L$ \ \|\ $Q_C$ & $Q_L$ & $Q_1$ &
$Q_2$
 & $Q_3$ & $Q_4$  \ \|\ $SU(5)$ $\times$ $SU(3)$ \ \|\ $Q_7$ &
$Q_8$  \crthick
$V_1$ \|\ ${b_1+2\beta}+(I)$ \|(1,1)\|~0 & ~~0 & ~~0 & ~~${1\over 2}$ &
 ~~$1\over 2$ & ~~0 \|(1,3)\| $-{1\over 2}$ &
{}~~$5\over 2$   \nr
$V_2$ \|\                \|(1,1)\| ~~0 & ~~0 & ~~0 & ~~${1\over 2}$ &
 ~~$1\over 2$ & ~~0  \|(1,$\bar 3$)\| ~~${1\over 2}$ &
$-{5\over 2}$  \nr
$V_3$ \|\                \|(1,1)\| ~~0 & ~~0 & ~~0 & ~~${1\over 2}$ &
 ~~$1\over 2$ & ~~0  \|(5,1)\| $-{1\over 2}$ &
$-{3\over 2}$  \nr
$V_4$ \|\                \|(1,1)\| ~~0 & ~~0 & ~~0 & ~~${1\over 2}$ &
 ~~$1\over 2$ & ~~0  \|($\bar 5$,1)\| ~~$1\over 2$ &
{}~~${3\over 2}$  \cr
$V_{5}$ \|\ ${b_2+2\beta}+(I)$ \|(1,1)\| ~~0 & ~~0 & ~~${1\over 2}$ & ~~0 &
 ~~$1\over 2$ & ~~0  \|(1,3)\| $-{1\over 2}$ &
 ~~$5\over 2$  \nr
$V_{6}$ \|\                \|(1,1)\| ~~0 & ~~0 & ~~${1\over 2}$ & ~~0 &
 ~~$1\over 2$ & ~~0  \|(1,$\bar 3$)\| ~~${1\over 2}$ &
 $-{5\over 2}$  \nr
$V_{7}$ \|\                \|(1,1)\| ~~0 & ~~0 & ~~${1\over 2}$ & ~~0 &
 ~~$1\over 2$ & ~~0  \|(5,1)\| $-{1\over 2}$ &
 $-{3\over 2}$ \nr
$V_{8}$ \|\                \|(1,1)\| ~~0 & ~~0 & ~~${1\over 2}$ & ~~0 &
 ~~$1\over 2$ & ~~0  \|($\bar 5$,1)\| ~~$1\over 2$ &
 ~~${3\over 2}$  \cr
$V_{9}$ \|\ ${b_3+2\beta}+(I)$ \|(1,1)\| ~~0 & ~~0 & ~~${1\over 2}$ &
 ~~${1\over 2}$ & ~~0 & ~$-{1\over 2}$  \|(1,3)\| $-{1\over 2}$ &
 ~~$5\over 2$  \nr
$V_{10}$ \|\                \|(1,1)\| ~~0 & ~~0 & ~~${1\over 2}$ &
 ~~${1\over 2}$ & ~~0 & ~$-{1\over 2}$
 \|(1,$\bar 3$)\| ~~${1\over 2}$
 & $-{5\over 2}$  \nr
$V_{11}$ \|\                \|(1,1)\| ~~0 & ~~0 & ~~${1\over 2}$ &
 ~~${1\over 2}$ & ~~0 & ~~${1\over 2}$  \|(5,1)\|
 $-{1\over 2}$ & $-{3\over 2}$  \nr
$V_{12}$ \|\                \|(1,1)\| ~~0 & ~~0 & ~~${1\over 2}$ &
 ~~${1\over 2}$ & ~~0 & ~~${1\over 2}$  \|($\bar 5$,1)\| ~~$1\over 2$
 & ~~${3\over 2}$
 \endtable}
\hfill}
\bigskip
\parindent=0pt
\hangindent=39pt\hangafter=1
{\it Table 6.} Massless states in model 5
and their quantum numbers. V indicates
that these states form vector representations of the Hidden group.

\vfill
\eject

\input tables.tex
\nopagenumbers
\magnification=1000
\baselineskip=18pt
\hbox
{\hfill
{\begintable
\ F \ \|\ SEC \ \|\ $SU(3)_C$ $\times$ $SU(2)_L$ \ \|\ $Q_C$ & $Q_L$ & $Q_1$ &
$Q_2$
 & $Q_3$ & $Q_4$  \ \|\ $SU(5)$ $\times$ $SU(3)$ \ \|\ $Q_7$ &
$Q_8$  \crthick
$M_1$ \|\ $S+{b_1}+{b_2}$ \|(3,1)\| ~~${1\over4}$ & ~~${1\over2}$
 & $-{1\over4}$ & ~~${1\over 4}$ &
 $-{1\over 4}$ & ~~0  \|(1,1)\| $-{1\over 4}$ &
$-{{15}\over 4}$ \nr
$M_2$ \|\ $+\beta+\gamma+(I)$ \|(1,2)\| ~~${3\over4}$ & ~~${1\over2}$
 & ~~${1\over4}$ & $-{1\over 4}$ &
 ~~${1\over 4}$ & ~~0  \|(1,1)\| ~~${1\over 4}$ &
{}~~${{15}\over 4}$  \nr
$M_3$ \|\                \|(1,1)\| $-{3\over4}$ & ~~${1\over2}$
 & ~~${3\over4}$ & ~~${1\over 4}$ &
 $-{1\over 4}$ & ~~0  \|(1,1)\| $-{1\over 4}$ &
${-{15}\over 4}$ \nr
$M_4$ \|\                \|(1,1)\| ~~${3\over4}$ & $-{1\over2}$
 & ~~${1\over4}$ & ~~${3\over 4}$ &
 ~~${1\over 4}$ & ~~0 \|(1,1)\| ~~${1\over 4}$ &
{}~~${{15}\over 4}$  \nr
$M_5$ \|\                \|(1,1)\|$-{3\over4}$ & ~~$1\over2$
 & ~~${1\over4}$ & ~~${1\over 4}$ &
 ~~${3\over 4}$ & ~~0  \|(1,1)\| $-{1\over 4}$ &
${-{15}\over 4}$   \nr
$M_6$ \|\                \|(1,1)\| $-{3\over4}$ & ~~${1\over2}$
& $-{1\over4}$ & ~~${1\over 4}$ &
 $-{1\over 4}$ & ~~0
\|(5,1)\| $-{1\over 4}$ & ~~${{9}\over 4}$  \nr
$M_7$ \|\                \|(1,1)\| ~~${3\over4}$ & $-{1\over2}$
 & ~~${1\over4}$ & $-{1\over 4}$ &
 ~~${1\over 4}$ & ~~0  \|(1,3)\| $-{3\over 4}$ &
$-{5\over 4}$ \cr
$M_8$ \|\ $S+{b_1}+{b_2}$ \|(${\bar 3}$,1)\| $-{1\over4}$
& $-{1\over2}$
 & $-{1\over4}$ & ~~${1\over 4}$ &
 ~~${1\over 4}$ & ~~0  \|(1,1)\| ~~$1\over 4$ &
{}~~${{15}\over 4}$ \nr
$M_9$ \|\ $+\alpha{\pm\gamma}+(I)$ \|(1,2)\| $-{3\over4}$ & $-{1\over2}$
 & ~~${1\over4}$ & $-{1\over 4}$ &
 $-{1\over 4}$ & ~~0  \|(1,1)\| $-{1\over 4}$ &
$-{{15}\over 4}$  \nr
$M_{10}$ \|\                \|(1,1)\| $-{3\over4}$ & ~~${1\over2}$
 & $-{3\over4}$ & $-{1\over 4}$ &
 $-{1\over 4}$ & ~~0  \|(1,1)\| $-{1\over 4}$ &
$-{{15}\over 4}$  \nr
$M_{11}$ \|\                \|(1,1)\| ~~${3\over4}$ & $-{1\over2}$
 & $-{1\over4}$ & $-{3\over 4}$ &
 ~~${1\over 4}$ & ~~0
\|(1,1)\| ~~${1\over 4}$ &
{}~~${{15}\over 4}$  \nr
$M_{12}$ \|\                \|(1,1)\| ~~${3\over4}$ & $-{1\over2}$
 & $-{1\over4}$ & ~~${1\over 4}$ &
 $-{3\over 4}$ & ~~0
\|(1,1)\| ~~${1\over 4}$ &
{}~~${{15}\over 4}$ \nr
$M_{13}$ \|\         \|(1,1)\| ~~${3\over4}$ & $-{1\over2}$
 & $-{1\over4}$ & ~~${1\over 4}$ &
 ~~${1\over 4}$ & ~~0
\|(${\bar 5}$,1)\| ~~${1\over 4}$ &
$-{{9}\over 4}$  \nr
$M_{14}$ \|\                \|(1,1)\| $-{3\over4}$ & ~~${1\over2}$
 & ~~${1\over4}$ & $-{1\over 4}$ &
 $-{1\over 4}$ & ~~0  \|(1,${\bar 3}$)\| ~~$3\over 4$ &
{}~~${5\over 4}$  \cr
$M_{15}$ \|\ $1+{b_1}+{b_2}+{b_3}$  \|(1,2)\| ~~0 & ~~0
 & $-{1\over2}$ & ~~0 &
 ~~0 & ~~${1\over2}$
\|(1,1)\| ~~1 & ~~0  \nr
$M_{16}$ \|\ $+\beta+2\gamma$ \|(1,2)\| ~~0 & ~~0
 & $-{1\over2}$ & ~~0 &
 ~~0 & $-{1\over2}$
\|(1,1)\| ~~1 & ~~0  \nr
$M_{17}$ \|\                  \|(1,1)\| ~~0 & ${-1}$
 & ~~${1\over2}$ & ~~0 &
 ~~0 & ~~${1\over2}$
\|(1,1)\| ~~1 & ~~0  \nr
$M_{18}$ \|\                   \|(1,1)\| ~~0 & ${-1}$
 & ~~${1\over2}$ & ~~0 &
 ~~0 & $-{1\over2}$
\|(1,1)\| ~~1 & ~~0  \nr
$M_{19}$ \|\                   \|(1,1)\| ~~0 & ${-1}$
 & $-{1\over2}$ & ~~0 &
 ~~0 & $-{1\over2}$
\|(1,1)\| $-1$ & ~~0  \nr
$M_{20}$ \|\                   \|(1,1)\| ~~0 & ${-1}$
 & $-{1\over2}$ & ~~0 &
 ~~0 & ~~${1\over2}$
\|(1,1)\| $-1$ & ~~0  \cr
$M_{21}$ \|\ $1+{b_1}+{b_2}+{b_3}$ \|(1,2)\| ~~0 & ~~0
& ~~0 & ~~${1\over2}$  &
 ~~0 & ~~${1\over2}$
\|(1,1)\| $-1$ & ~~0  \nr
$M_{22}$ \|\ $+\alpha+2\gamma$  \|(1,2)\| ~~0 & ~~0
& ~~0 & ~~${1\over2}$  &
 ~~0 & $-{1\over2}$
\|(1,1)\| $-1$ & ~~0  \nr
$M_{23}$ \|\                    \|(1,1)\| ~~0 & ~~${1}$
 & ~~0 & $-{1\over 2}$ &
 ~~0 & ~~${1\over2}$
\|(1,1)\| $-1$ & ~~0  \nr
$M_{24}$ \|\                    \|(1,1)\| ~~0 & ~~${1}$
& ~~0 & $-{1\over2}$  &
 ~~0 & $-{1\over2}$
\|(1,1)\| $-1$ & ~~0 \nr
$M_{25}$ \|\                    \|(1,1)\| ~~0 & ~~${1}$
& ~~0 & ~~${1\over2}$  &
 ~~0 & $-{1\over2}$
\|(1,1)\| ~~1 & ~~0  \nr
$M_{26}$ \|\                     \|(1,1)\| ~~0 & ~~${1}$
& ~~0 & ~~${1\over2}$  &
 ~~0 & ~~${1\over2}$
\|(1,1)\| ~~1 & ~~0
\endtable}
\hfill}
\bigskip
\parindent=0pt
\hangindent=39pt\hangafter=1
{\it Table 7.} Massless states in model 5 and their quantum numbers.

\end
\bye